\documentclass[pdflatex,sn-mathphys-num]{sn-jnl} % Numbered (Vancouver) style

\usepackage{tabularx}
\usepackage{graphicx}
\usepackage{multirow}
\usepackage{amsmath,amssymb,amsfonts}
\usepackage{booktabs}
\usepackage{xcolor}
\usepackage{hyperref}
\usepackage{pgfplots}
\pgfplotsset{compat=1.18}
\usepackage{amsthm}

\usepackage[title]{appendix}
\usepackage{textcomp}
\usepackage{manyfoot}
\usepackage{algorithm}
\usepackage{algorithmicx}
\usepackage{algpseudocode}
\usepackage{listings}
\usepackage{array}      % <-- m{...} p{...} column types
\usepackage{float}
\let\mathscr\undefined
\usepackage[mathscr]{euscript}
\theoremstyle{thmstyleone}

\theoremstyle{thmstyletwo}

\theoremstyle{thmstylethree}

\begin{document}

%%%%% TITLE %%%%%
\title[ROI in Chronic Disease Management]{Modeling ROI in chronic disease management: a simulation-based framework integrating patient adherence and policy timing}

%%%%% AUTHORS %%%%%
\author*[1]{\fnm{Jinho} \sur{Cha}}\email{jcha@GwinnettTech.edu}
\author[2]{\fnm{Eunchan Daniel} \sur{Cha}}
\author[3]{\fnm{Emily Haneul} \sur{Yoo}}
\author[4]{\fnm{Hyoshin} \sur{Song}}

\affil*[1]{\orgdiv{Department of Computer Science}, \orgname{Gwinnett Technical College}, \city{Georgia}, \country{USA}}
\affil[2]{\orgdiv{School of Biological Sciences}, \orgname{Georgia Institute of Technology}, \city{Georgia}, \country{USA}}
\affil[3]{\orgname{North Gwinnett High School}, \city{Suwanee}, \country{USA}}
\affil[4]{\orgname{Oakton High School}, \city{Vienna}, \country{USA}}

%%%%% ABSTRACT %%%%%
\abstract{
\textbf{Background:} Chronic diseases impose a sustained burden on healthcare systems through progressive deterioration and long-term costs. Although adherence-enhancing interventions are widely promoted, their return on investment (ROI) remains uncertain, particularly under heterogeneous patient behavior and socioeconomic variation.

\textbf{Methods:} We developed a simulation-based framework integrating disease progression, time-varying adherence, and policy timing. Cumulative healthcare costs were modeled over a 10-year horizon using continuous-time stochastic formulations calibrated with Medical Expenditure Panel Survey (MEPS) data stratified by income. ROI was estimated across adherence gains ($\delta$) and policy costs ($\gamma$).

\textbf{Results:} Early and adaptive interventions yielded the highest ROI by sustaining adherence and slowing progression. ROI exceeded 20\% when $\delta \geq 0.20$ and $\gamma \leq 1.5$, whereas low-impact or high-cost policies failed to break even. Subgroup analyses showed a 32\% ROI gap between the lowest and highest income strata, with projected savings of \$312 per patient versus baseline. Sensitivity tests confirmed robustness under stochastic adherence and inflation variability.

\textbf{Conclusions:} The framework offers a transparent, adaptable tool for evaluating cost-effective adherence strategies. By linking behavioral effectiveness with fiscal feasibility, it supports design of robust and equitable chronic disease policies. Reported ROI values represent conservative lower bounds, and extensions incorporating DALYs and QALYs illustrate scalability toward health outcome integration.
}

%%%%% KEYWORDS %%%%%

\maketitle

\section{Introduction}\label{sec1}

Chronic diseases—such as diabetes, cardiovascular disorders, and chronic respiratory conditions—account for more than 70\% of global deaths annually, disproportionately affecting low-income and aging populations \cite{who2018non}. In the United States, chronic care represents over 85\% of total health expenditure \cite{cdc2022costs}. As this burden intensifies, developing efficient and equitable strategies for resource allocation has become a central challenge in sustainable public health management.

Traditional cost-modeling approaches, including actuarial tables, regression-based risk models, and Markov chains, have long guided expenditure projections and risk forecasting \cite{thorpe2004chronic,zhang2017long,taylor2013chronic,russell2009costeffectiveness}. While valuable for population-level planning, these models rely on static covariates and linear assumptions that overlook dynamic feedback among disease progression, behavioral adherence, and policy timing, limiting their ability to capture complex real-world interactions.

Recent advances in simulation-based modeling address this gap by incorporating system dynamics, feedback loops, and behavioral economics into public health evaluation \cite{milstein2011health,best2012simulation,hunter2017system}. Yet most frameworks still analyze components—such as disease severity \cite{tappenden2004interferon}, patient adherence \cite{cutler2010value}, or behavioral nudges \cite{thaler2008nudge}—in isolation, with stochastic variation in patient behavior often underrepresented \cite{berwick2008triple}. 

The economic and clinical burden of chronic diseases has been widely studied across diverse modeling paradigms. Time-series, multivariate regression, and actuarial models remain standard for expenditure forecasting \cite{milstein2011health,hunter2017nhs}, but lack adaptability to evolving behavioral and policy contexts. Simulation approaches such as Markov models, system dynamics, and agent-based models (ABMs) address these limitations by representing nonlinear interactions, stochasticity, and feedback \cite{yang2013economic,freeman2019abm,elsayed2012social,galea2010causal}. ABMs can further embed social norms, adherence rules, and local interactions absent in conventional models \cite{auchincloss2008dynamic,gonzalezparra2012influenza,taylor2013chronic,russell2009costeffectiveness}. Hybrid frameworks combining system dynamics and ABMs enable real-time representation of service delivery, disease trajectories, and cost propagation \cite{atkinson2015policy,lee2025modeling,thorpe2004chronic,zhang2017long}. Despite such progress, few studies integrate patient-level feedback or temporally activated policy levers to assess ROI over long horizons \cite{sullivan2021equity,gupta2025roi}, and most cost-effectiveness analyses remain outcome-centric—focused on QALY or DALY metrics—rather than adherence-driven ROI dynamics.

Parallel developments in machine learning (ML) and predictive analytics have enhanced healthcare cost modeling. Using clinical, demographic, and behavioral data from electronic health records and claims databases, ML models can accurately forecast expenditures \cite{tran2025deep,patel2025adherence,nguyen2025mental,xu2021nhanes,hoehn2021ml}. They identify key cost drivers such as depression, LDL levels, and comorbidities, while informing risk-based policy design—e.g., prioritizing adherence reminders for high-risk patients \cite{chen2023frailty,rasheed2022shap}. Yet these models are primarily predictive rather than mechanistic, limiting transparency and scenario comparability for policy evaluation.

Adherence remains central to both health outcomes and healthcare costs. Poor medication adherence increases hospitalizations and mortality \cite{ho2006adherence,sokol2005adherence}, whereas modest improvements yield substantial long-term savings \cite{cutler2018economic}. However, most adherence models assume uniform behavioral responses, overlooking heterogeneity linked to socioeconomic status, mental health, and provider engagement \cite{kane2013nonadherence,zullig2015adherence}. Such behavioral variation underscores the need for frameworks that explicitly model heterogeneity and temporal adherence dynamics.

Emerging policy simulation models have begun to explore the timing and persistence of intervention effects. Microsimulation and dynamic cohort studies show that early behavioral interventions can markedly reduce downstream costs, particularly in high-risk subgroups \cite{vanbaal2016future,bilinski2020covid}, whereas delayed or generic programs often yield diminishing returns \cite{kim2020perspective}. Adaptive policy approaches that trigger interventions based on evolving risk or adherence trajectories have been proposed to improve efficiency \cite{choi2023adaptive}. These trends reflect a broader movement toward precision health policy—linking individual behavioral data to population-scale economic evaluation.

Nonetheless, key research gaps remain. Few frameworks jointly model disease progression, adherence dynamics, and policy activation within a unified simulation structure. Continuous-time formulations enabling flexible scenario exploration are rare, and stochastic behavioral variation is often simplified. Moreover, integration of nationally representative datasets such as MEPS and NHANES remains limited \cite{meps2023survey,nhanes2022survey}. Addressing these limitations requires models that are mathematically tractable, interpretable, and empirically grounded.

Building on recent advances in health economics and behavioral simulation \cite{simonsen2024intervention,evans2021financing,gupta2025roi}, we propose a modular, ROI-focused framework that integrates adherence dynamics, disease progression, and policy timing into a continuous-time simulation calibrated to empirical data. The model links behavioral economics and healthcare-cost dynamics to yield interpretable, policy-relevant insights.

Our contributions are threefold. First, we present a unified framework bridging behavioral economics with health-cost modeling. Second, we identify nonlinear ROI thresholds to optimize intervention timing and intensity. Third, we demonstrate empirical validity and adaptability across income-stratified populations using MEPS and NHANES data. The remainder of this paper outlines the model formulation (Section~\ref{sec:model_formulation}), intervention scenarios and simulation results (Sections~\ref{sec:intervention_scenarios}--\ref{sec:results}), and policy implications (Section~\ref{sec:discussion}). Extended derivations, robustness tests, and parameter tables are provided in the Supplementary Appendices.

%%%%%%%%%%%%%%%%%%%%
\section{Model formulation}
\label{sec:model_formulation}

This section introduces the core structure of our policy evaluation model, integrating disease progression, behavioral adaptation, and policy cost into a unified cost function. All subsequent intervention scenarios in Section~\ref{sec:intervention_scenarios} are grounded in this mathematical framework.

%---------------------
\subsection{Model structure: cumulative cost function}

Cumulative cost was defined as the discounted sum of instantaneous costs over time,  
\(C(t)=C_0+\int_{0}^{t} e^{-\rho s} c(s)\,ds\), where $\rho$ is the continuous discount rate.  
Instantaneous cost $c(s)$ combines four components—disease burden, adherence effort, policy expenditure, and an optional health–outcome term:

\begin{equation}
c(s)=
\frac{\alpha D_{\max}}{1 + e^{-k(s - s_0)}} +
\beta A(s)^2 +
\gamma P(s) +
\lambda H(s).
\label{eq:integrand}
\end{equation}

This compact formulation provides the structural basis for all policy scenarios.  
Detailed definitions and parameter values are listed in \textit{Supplementary Appendix E}.  
Figure~\ref{fig:S1_cost_decomp} illustrates the relative contribution of each component under an Early Adherence scenario.

\begin{figure}[htbp]
  \centering
  \includegraphics[width=0.7\linewidth]{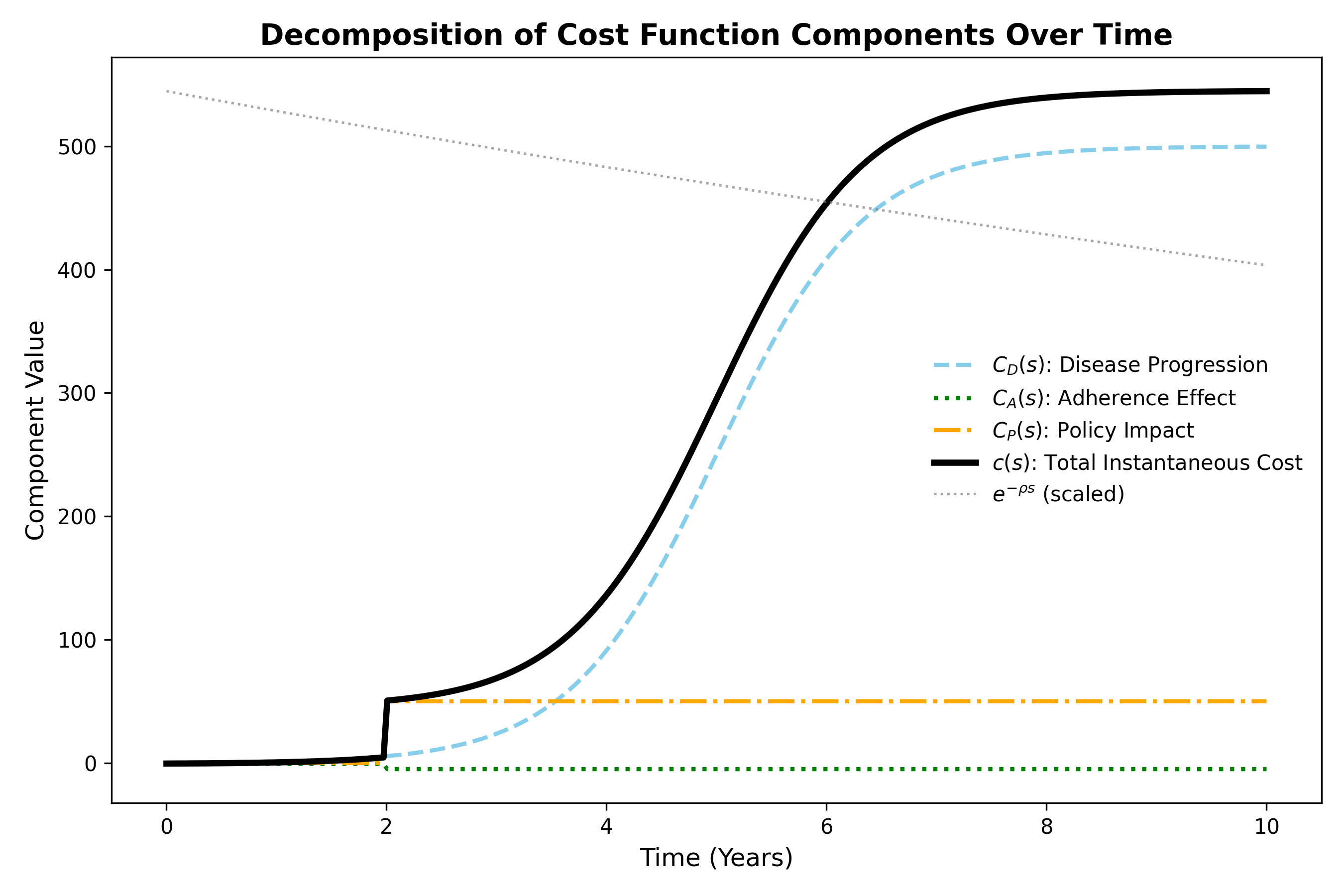}
  \caption{Instantaneous cost $c(s)$ decomposition under an Early Adherence scenario.
  Total cost (black) is composed of disease burden (blue), adherence term (green),
  and policy expenditure (orange). The dotted line indicates the discount factor $e^{-\rho s}$ (scaled).}
  \label{fig:S1_cost_decomp}
\end{figure}

\subsection{Parameter definitions and calibration}

Table~\ref{tab:params} lists key model parameters, their roles, and data sources. 
Parameters were estimated using MEPS and NHANES data (2015–2023), with policy costs benchmarked to CDC program reports.

\begin{table}[htbp]
\centering
\scriptsize
\renewcommand{\arraystretch}{1.2}
\caption{Model parameters and empirical sources.}
\label{tab:params}
\begin{tabularx}{\textwidth}{lXl}
\toprule
\textbf{Parameter} & \textbf{Definition} & \textbf{Source / Notes} \\
\midrule
$C_0$ & Baseline annual cost per patient & MEPS 2015–2023 \\
$\rho$ & Discount rate (3\%) & US Panel on Cost-Effectiveness \\
$D_{\max},k,s_0$ & Disease progression parameters & NHANES prevalence curves \\
$\alpha$ & Disease cost multiplier & MEPS calibration \\
$A_0,\delta$ & Baseline adherence and policy shift & NHANES medication-use data \\
$\tau$ & Policy start time & Intervention year \\
$\beta$ & Adherence cost weight ($\beta<0$ indicates savings) & Calibrated value \\
$P(s),\gamma$ & Policy cost function and scale & CDC/NIH program benchmarks \\
\bottomrule
\end{tabularx}
\end{table}

\subsection{Stochastic behavior modeling}

Adherence gain $\delta$ was modeled as a random variable from a Beta distribution fitted to MEPS adherence data.  
Each simulation drew $\delta$ and propagated it through the cost model to obtain distributions of total cost and ROI.  
A binary alternative (high vs. low responders, probability $p$) was tested using NHANES subgroup prevalence.

\subsection{Simulation design}

Simulations were run in Python (NumPy/SciPy) over a 10-year horizon.  
Inputs were drawn from MEPS (cost, adherence) and NHANES (disease burden, comorbidity).  
Four cases were compared: (i) no policy, (ii) deterministic adherence shift, (iii) stochastic Beta uptake, and (iv) subgroup heterogeneity.  
Outcomes included cumulative cost $C(t)$, ROI, and payback time (years to net savings).

\subsection{Scenario integration}

All intervention scenarios (Section~\ref{sec:intervention_scenarios}) use the same structural model, differing only in policy timing ($\tau$), adherence response ($\delta$), and cost intensity ($\gamma, P(s)$).  
This uniform design ensures that observed differences in outcomes reflect policy configuration rather than model specification.

%===============================
\section{Intervention Scenarios}
\label{sec:intervention_scenarios}
%===============================

To assess the cost-effectiveness and behavioral dynamics of chronic care policies, 
we simulated six policy scenarios differing in intervention timing, adherence response, 
and policy cost intensity (Table~\ref{tab:scenario_settings}). 
These configurations span realistic contrasts—from proactive early engagement to inefficient, high-cost programs—
and were calibrated using MEPS and NHANES data on chronic disease and treatment adherence
\cite{meps2023survey, nhanes2022survey}.

All scenarios share a unified cost framework:
\begin{equation}
c_i(s) =
\frac{\alpha D_{\max}}{1 + e^{-k(s - s_0)}} +
\beta A_i(s)^2 +
\gamma_i P_i(s),
\label{eq:scenario_general}
\end{equation}
where $A_i(s)$ represents adherence behavior and $P_i(s)$ denotes policy expenditure for scenario $i$.  
Cumulative cost is obtained as \(C(t) = C_0 + \int_0^t e^{-\rho s} c_i(s)\,ds\).  
Economic performance was summarized by the return on investment (ROI),  
defined as ROI = \((C_{\text{baseline}} - C_{\text{policy}})/C_{\text{policy}}\).

\begin{table}[htbp]
\centering
\small
\setlength{\tabcolsep}{3pt}
\renewcommand{\arraystretch}{1.2}
\caption{Policy scenario configurations.}
\label{tab:scenario_settings}
\begin{tabular}{lccc p{4.3cm}}
\toprule
\textbf{Scenario} & \textbf{Start ($\tau$)} & \textbf{Adherence $\Delta\delta$} & \textbf{Cost ($\gamma$)} & \textbf{Description} \\
\midrule
Baseline & -- & 0 & 0 & Natural progression; reference benchmark. \\
Early Adherence & 2 & +0.3 & 1.5 & Early proactive engagement. \\
Delayed Intervention & 5 & +0.3 & 1.5 & Late reactive response. \\
Regressive Adherence & 2 & +0.3→0 & 1.2 & Behavioral decay over time. \\
Adaptive Nudges & varies & +0.3 (stepwise) & 2.0 & Dynamic re-engagement to sustain adherence. \\
Low-Impact Policy & 2 & +0.05 & 3.0 & High spending, minimal behavioral change. \\
\bottomrule
\end{tabular}
\vspace{3pt}
\footnotesize\textit{Parameter values are illustrative, calibrated to MEPS/NHANES (2015–2023).}
\end{table}

These six scenarios form the basis for the simulation results in Section~\ref{sec:results}, 
allowing outcome differences to be directly attributed to policy design rather than model specification.  
Detailed scenario-specific cost formulations (\(c_i(s)\)) and adherence trajectories 
are provided in \textit{Supplementary Appendix E} for full mathematical reference.

---

\subsection{Scenario 1: Baseline (No Intervention)}

In the absence of intervention, adherence remains constant \(A(s)=A_0\) and policy cost is zero \(P(s)=0\).  
This represents natural disease progression and serves as the counterfactual benchmark for ROI comparisons.

---

\subsection{Scenario 2: Early Adherence Activation}

An early intervention begins at \(\tau=2\) years, producing a discrete increase in adherence  
\(A(s)=A_0+\delta I(s-\tau)\) with cost \(P(s)=I(s-\tau)\).  
This captures early-stage programs such as digital reminders or education campaigns.  
By improving adherence before major disease progression, early activation achieves sustained cost reduction over time.

---

\subsection{Scenario 3: Delayed Intervention}

A reactive policy starts later (\(\tau=5\)), using the same adherence shift \(A(s)=A_0+\delta I(s-5)\).  
Because improvement begins after disease burden has accumulated, total savings are limited.  
This highlights the diminishing benefit of delayed responses in chronic-care management.

---

\subsection{Scenario 4: Regressive Adherence}

Adherence initially rises but decays exponentially without reinforcement,  
\(A(s)=A_0+\delta e^{-\theta (s-\tau)} I(s-\tau)\).  
One-time incentives or static nudges yield short-lived gains, and both cost savings and health improvements erode as behavior returns toward baseline.

---

\subsection{Scenario 5: Adaptive Nudges}

This model introduces feedback-based re-engagement.  
Whenever adherence drops below a threshold \(A_{\text{thresh}}\), additional nudges are activated, increasing \(A(s)\) and incurring extra policy cost proportional to the number of activations.  
Appropriate threshold calibration maintains stable adherence and positive ROI, 
demonstrating that adaptive strategies can achieve long-term efficiency if properly tuned.

---

\subsection{Scenario 6: High-Cost, Low-Impact Policy}

Here, adherence gains are minimal \((\delta \ll 1)\) despite large expenditures \((\gamma \gg 1)\).  
\(A(s)=A_0+\delta I(s-\tau)\) and \(P(s)=I(s-\tau)\) represent an inefficient policy with weak behavioral change.  
Such imbalance between cost intensity and behavioral efficacy results in negative ROI and poor cost-effectiveness.

---

\subsection{Summary Insight}

Across all simulations, early or adaptive interventions yield the strongest and most sustainable cost-effectiveness, 
while delayed, decaying, or inefficient policies show limited or negative returns.  
These findings emphasize that \textit{timing, sustained engagement, and economic balance} 
are key to maximizing policy value in chronic care systems.

\section{Results}
\label{sec:results}

All results presented here are based on ROI defined in terms of direct healthcare 
expenditures ($\lambda = 0$ in Equation~\ref{eq:integrand}). Broader outcomes such as 
QALYs, DALYs, or societal productivity effects were not part of the primary analysis. 
Exploratory extensions with nonzero $\lambda$ (Sections~\ref{sec:results}, {Supplementary Appendix E}) 
provide illustrative examples, but a full integration of health-adjusted outcomes 
remains beyond the scope of this study.

\subsection{Scenario-based simulation outcomes}

The simulation model produced distinct outcomes across the baseline and six intervention scenarios, revealing clear trade-offs in clinical progression, adherence behavior, and long-term costs over a 10-year horizon.

\subsubsection{Baseline and intervention comparisons}

\begin{figure}[H]
    \centering
    \includegraphics[width=0.75\linewidth]{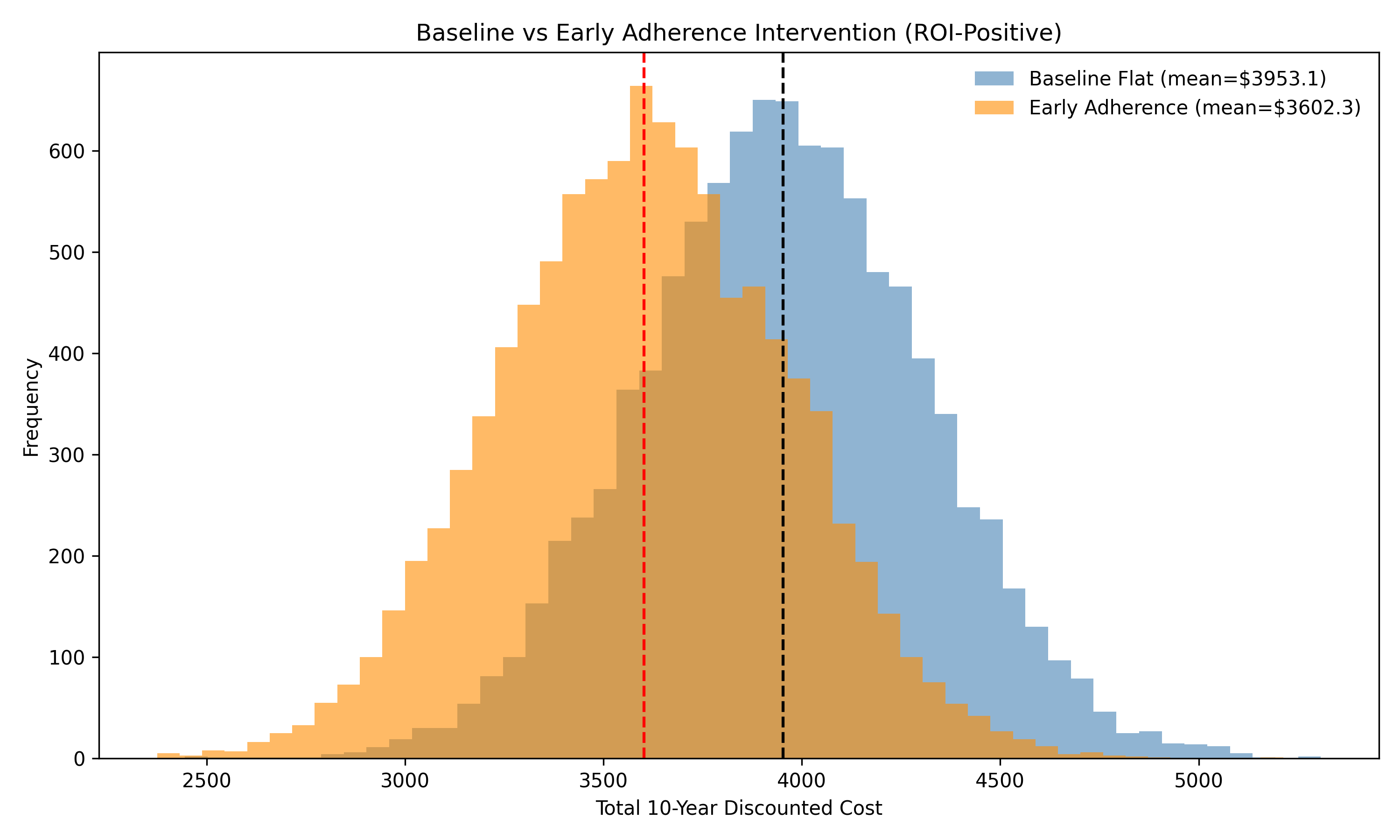}
    \caption{Baseline vs.\ Early Adherence intervention: Distribution of 10-year discounted healthcare costs across 10{,}000 Monte Carlo simulations. The baseline yielded a mean cost of \$3953.07, while the Early Adherence case reduced this to \$3602.26 (ROI = 9.7\%).}
    \label{fig:baseline_vs_early_final}
\end{figure}

The flat baseline scenario produced a mean 10-year discounted cost of \$3953.07, whereas the Early Adherence intervention reduced this to \$3602.26, yielding an average ROI of 9.7\%. Figure~\ref{fig:baseline_vs_early_final} illustrates clear divergence between baseline and intervention trajectories, confirming that early engagement achieves substantial long-term savings.

ROI in this study reflects direct healthcare expenditures only; broader effects such as avoided hospitalizations, QALYs, DALYs, or productivity gains were excluded. Accordingly, all reported ROI values should be viewed as conservative lower bounds of intervention value.

Policy interventions shifted these cost trajectories through adherence dynamics. Early and adaptive strategies provided the greatest cost containment and disease mitigation, whereas delayed, regressive, and low-impact approaches remained near baseline, indicating limited effectiveness.

%--------------------------
\subsubsection{Disease severity and adherence trajectories}

\begin{figure}[H]
    \centering
    \includegraphics[width=0.8\linewidth]{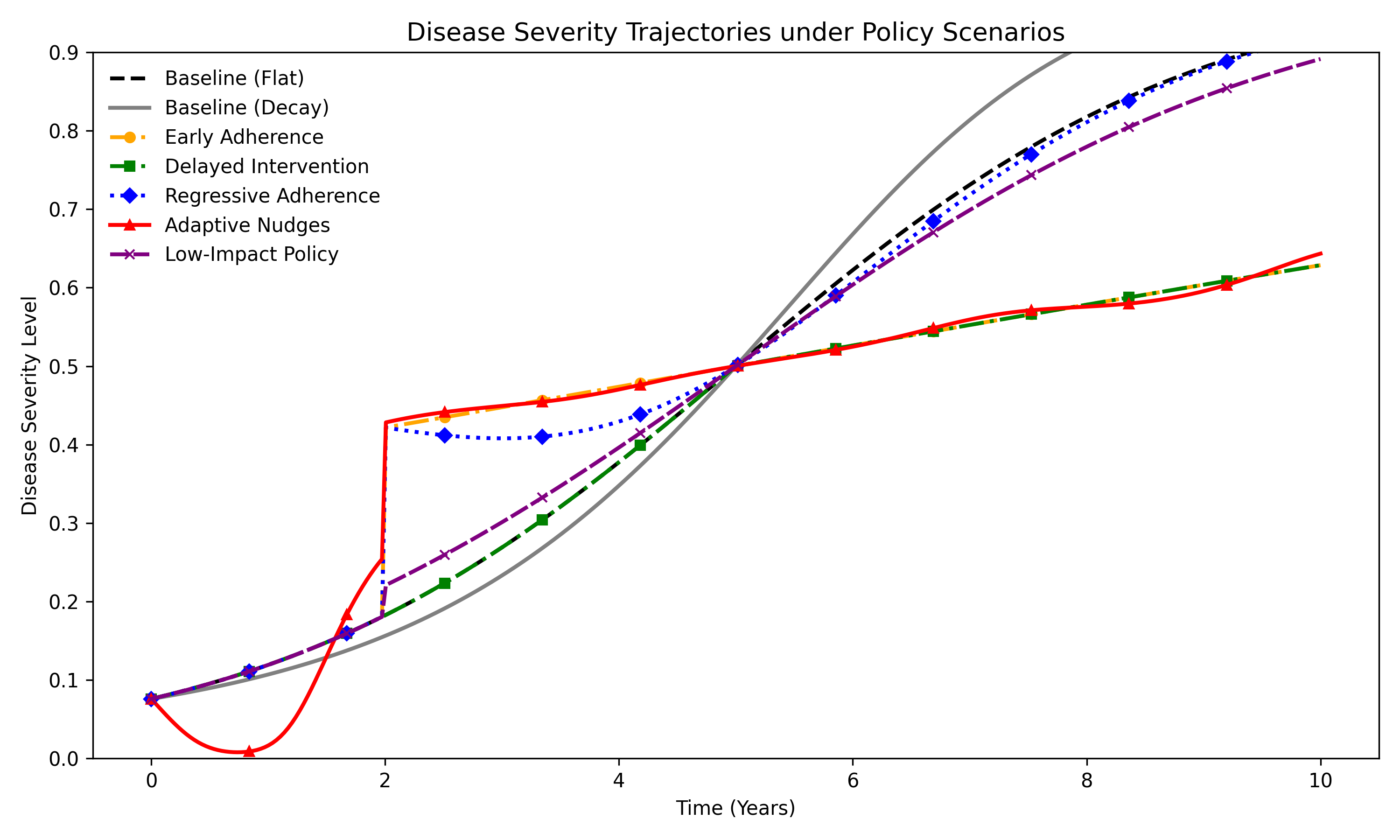}
    \caption{Disease severity trajectories under policy scenarios with flat vs. decaying baseline adherence. 
    Early Adherence and Adaptive Nudges lower final severity below 0.70, while Delayed, 
    Regressive, and Low-Impact interventions show minimal change. 
    The decay baseline illustrates that constant-adherence assumptions may inflate incremental benefits.}
    \label{fig:disease_progression_flat_vs_decay}
\end{figure}

Clinical outcomes diverged substantially across scenarios (Figure~\ref{fig:disease_progression_flat_vs_decay}). 
Early or adaptive adherence improvements reduced year-10 severity below 0.70, 
whereas weak or unsustained engagement left severity above 0.87. 
Including a decaying baseline reveals that assuming constant adherence can overstate incremental benefits and highlights the need for realistic counterfactuals.

\begin{figure}[H]
    \centering
    \includegraphics[width=0.8\linewidth]{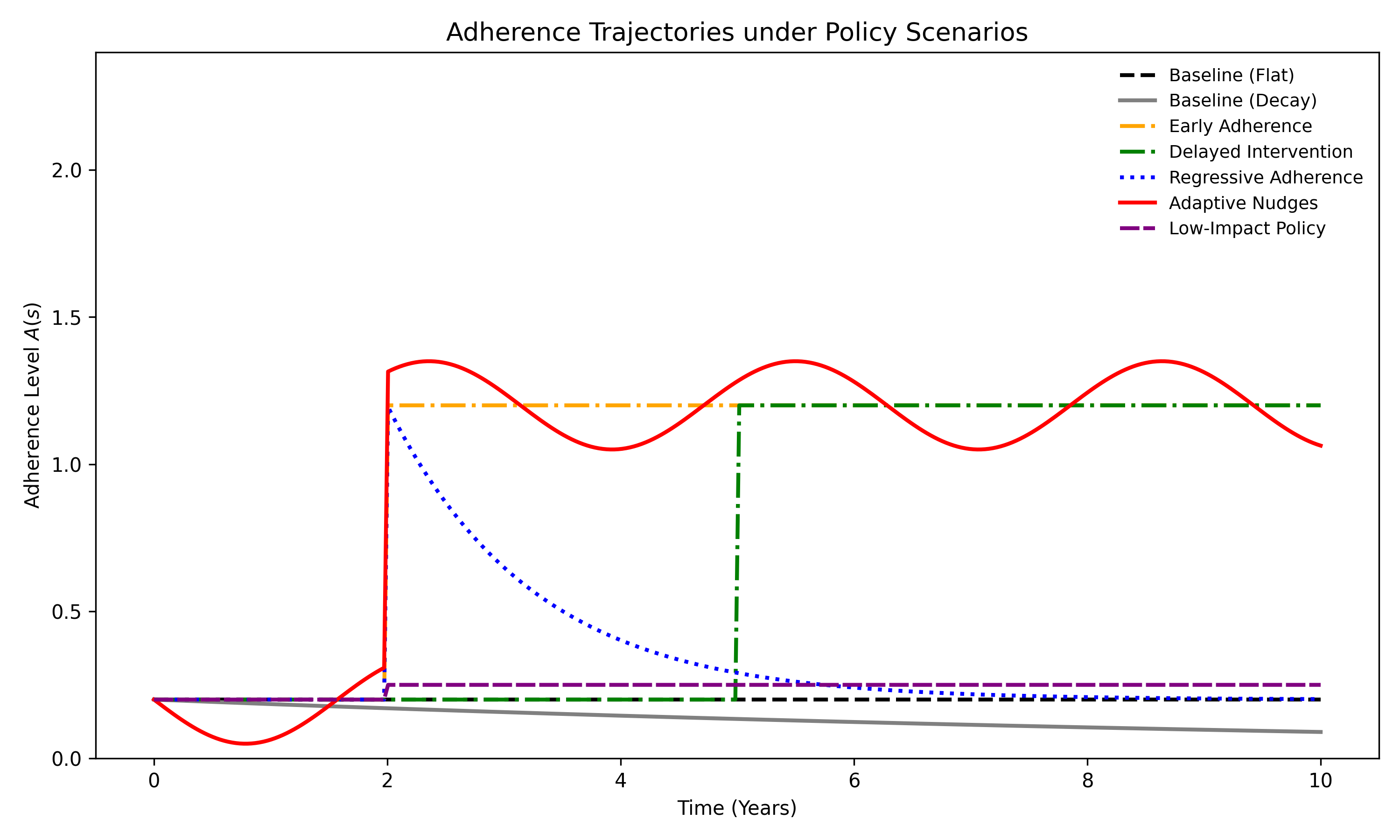}
    \caption{Adherence trajectories $A(s)$ across scenarios with flat vs. decaying baselines. 
    Early Adherence and Adaptive Nudges sustain adherence above 0.8, 
    while Regressive and Low-Impact designs decline toward baseline levels. 
    The decay baseline underscores that constant-adherence assumptions can exaggerate policy effects.}
    \label{fig:adherence_trend_flat_vs_decay}
\end{figure}

Differences in severity reflect adherence behavior over time (Figure~\ref{fig:adherence_trend_flat_vs_decay}). 
Early and adaptive policies maintained adherence above 0.8, supporting long-term control, 
whereas regressive and low-impact designs eroded quickly, dropping below 0.4 by year 10. 
Again, comparisons with a decaying baseline confirm that constant-adherence assumptions can exaggerate intervention value.

Overall, sustained adherence drives superior clinical outcomes: 
interventions fostering consistent engagement achieve better control, 
whereas late or decaying impacts yield little change. 
Both intervention timing and baseline choice therefore critically shape ROI estimates.
%--------------------------
\subsubsection{Cost trajectories across scenarios}

\begin{figure}[H]
    \centering
    \includegraphics[width=0.8\linewidth]{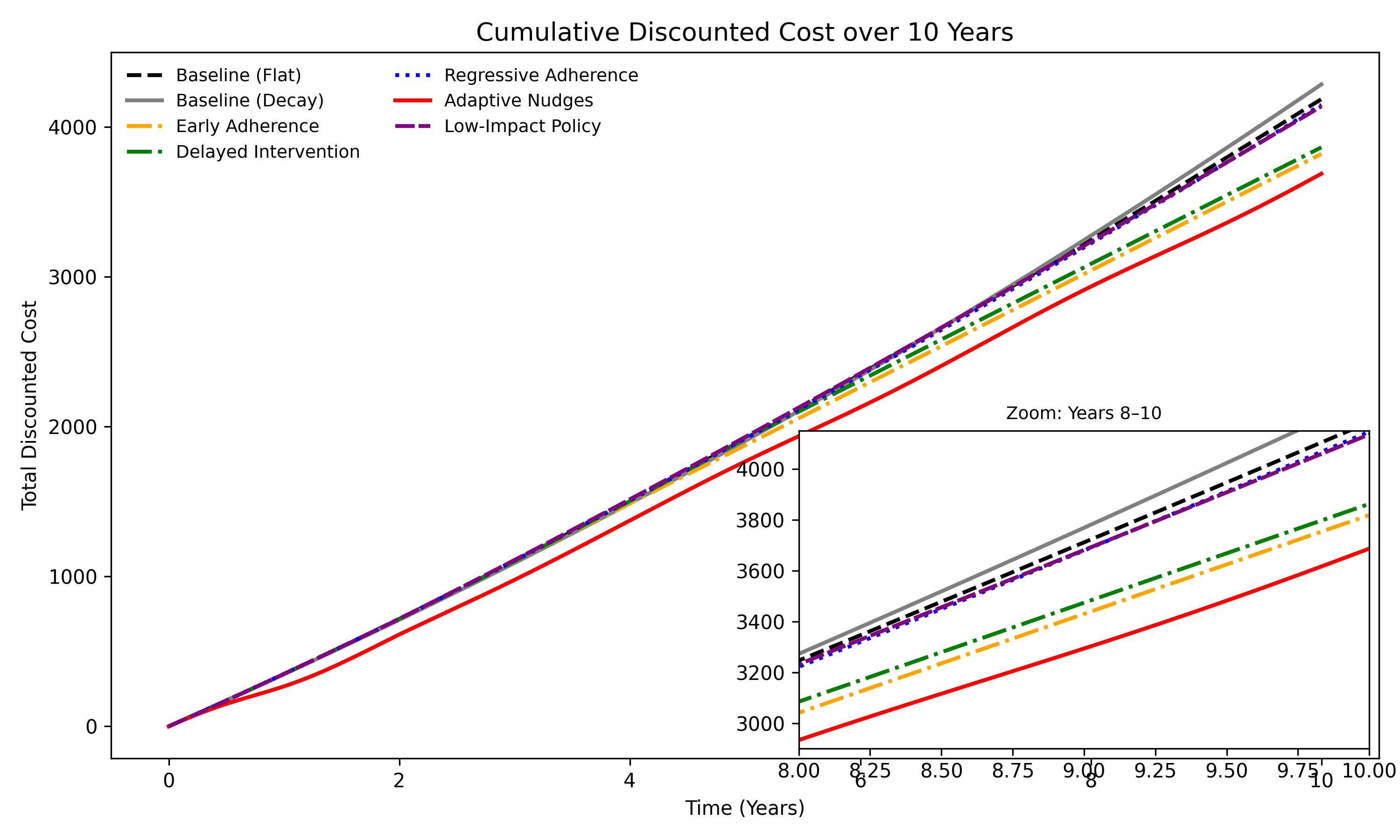}
    \caption{Ten-year cumulative discounted cost trajectories across scenarios, benchmarked 
    against both flat and decaying baselines. Early Adherence and Adaptive Nudges flatten 
    long-run costs to \$3900--4000 by year 10, while Delayed, Regressive, and Low-Impact 
    policies stay near or above baseline. Including a decaying baseline shows that constant 
    adherence may overstate savings. \textit{ROI values reflect direct healthcare expenditures only.}}
    \label{fig:cost_trajectory_flat_vs_decay}
\end{figure}

Figure~\ref{fig:cost_trajectory_flat_vs_decay} shows 10-year cumulative discounted 
cost trajectories across scenarios, benchmarked against flat and decaying baselines. 
Interventions with upfront investment—especially \textit{Early Adherence} and 
\textit{Adaptive Nudges}—incurred higher early costs but subsequently flattened 
expenditure growth, ending near \$3900--4000 by year 10. 
Thus, early, durable engagement mitigates clinical progression and prevents long-term 
cost escalation.

By contrast, the \textit{Low-Impact Policy} incurred the highest cumulative cost 
(>\$4600) despite minimal adherence or severity improvement. 
Delayed and regressive interventions produced intermediate results, driven by timing 
and adherence decay. Mapping these to ROI thresholds shows that cost containment 
depends on early, sustained adherence gains; weaker designs often fail to break even. 
Although ROI can be briefly positive in the Regressive scenario, discounting of future 
costs limits long-run benefit, underscoring the role of time-discounting in ROI evaluation.

Policies achieving strong, early adherence lower long-term costs, whereas late or weak 
interventions offer limited benefit. 
Baseline choice (flat vs. decay) further shows that conservative assumptions reduce 
incremental gains, reinforcing the need for comprehensive economic evaluation.

%------------------------------------------------------
\subsubsection{ROI landscape and cost-effectiveness frontier}

To assess the cost-effectiveness of adherence-based policies, 
return on investment (ROI) is defined as the relative cost savings over 
a 10-year horizon compared with the baseline (no intervention):

\begin{equation}
\text{ROI}(\delta, \gamma) = 
\left( \frac{C_{\text{baseline}} - C_{\text{policy}}(\delta, \gamma)}
{C_{\text{policy}}(\delta, \gamma)} \right) \times 100 ,
\label{eq:roi}
\end{equation}

where $C_{\text{baseline}}$ is the cumulative cost without policy and 
$C_{\text{policy}}(\delta, \gamma)$ the cost under a design defined by 
adherence gain $\delta$ and unit cost $\gamma$. 
Positive ROI denotes net savings; negative values indicate excess cost.

\begin{figure}[H]
    \centering
    \includegraphics[width=0.95\linewidth]{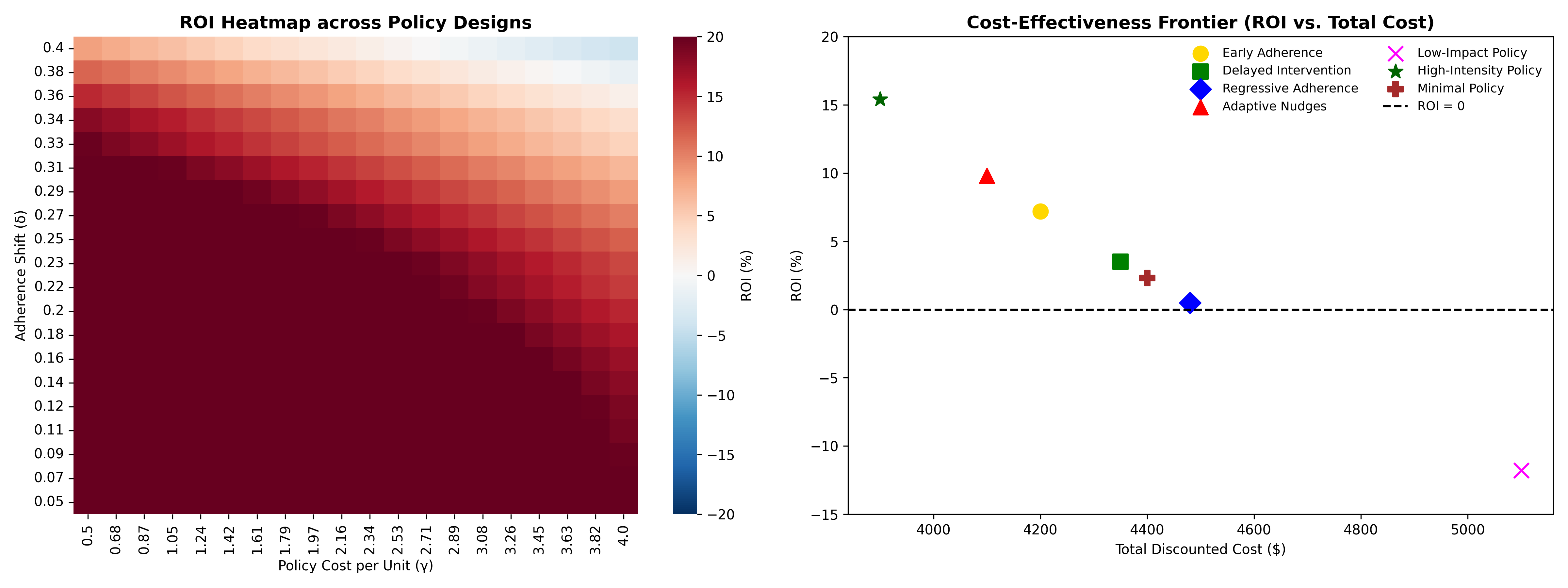}
    \caption{ROI landscape across policy designs. 
    Left: ROI heatmap over adherence gains ($\delta$) and policy costs ($\gamma$). 
    Right: Cost-effectiveness frontier showing ROI vs. total discounted cost. 
    Early Adherence and Adaptive Nudges remain ROI-positive, while costly or weak 
    designs fall below ROI = 0. \textit{ROI reflects direct healthcare expenditures only.}}
    \label{fig:roi_landscape}
\end{figure}

Figure~\ref{fig:roi_landscape} depicts the ROI design space. 
The left panel shows a heatmap of ROI values across $(\delta,\gamma)$, 
and the right panel the cost-effectiveness frontier relating ROI to total cost. 
Together they highlight trade-offs between behavioral effectiveness and cost intensity. 
\textit{Early Adherence} and \textit{Adaptive Nudges} occupy ROI-positive regions, 
whereas high-cost or low-impact designs fall below the break-even line.

To illustrate the extensibility of the $\lambda H(s)$ term, we examined how incorporating 
health-adjusted outcomes (e.g., DALYs or QALYs) affects ROI estimates. 
Assuming that each avoided DALY is valued at \$50{,}000, and that the 
\textit{Early Adherence} policy averts 0.05 DALYs per patient 
(corresponding to an additional monetized benefit of approximately \$2{,}500), 
the resulting ROI increases from 9.7\% to roughly 15\%. 
This simplified example demonstrates how translating health gains into fiscal terms 
can substantially enhance perceived economic value. 
In the present study, however, we set $\lambda = 0$, restricting ROI to direct healthcare 
expenditures. A conservative QALY-based sensitivity example is provided in 
\textit{Supplementary Appendix B}.

\begin{figure}[H]
    \centering
    \includegraphics[width=0.95\linewidth]{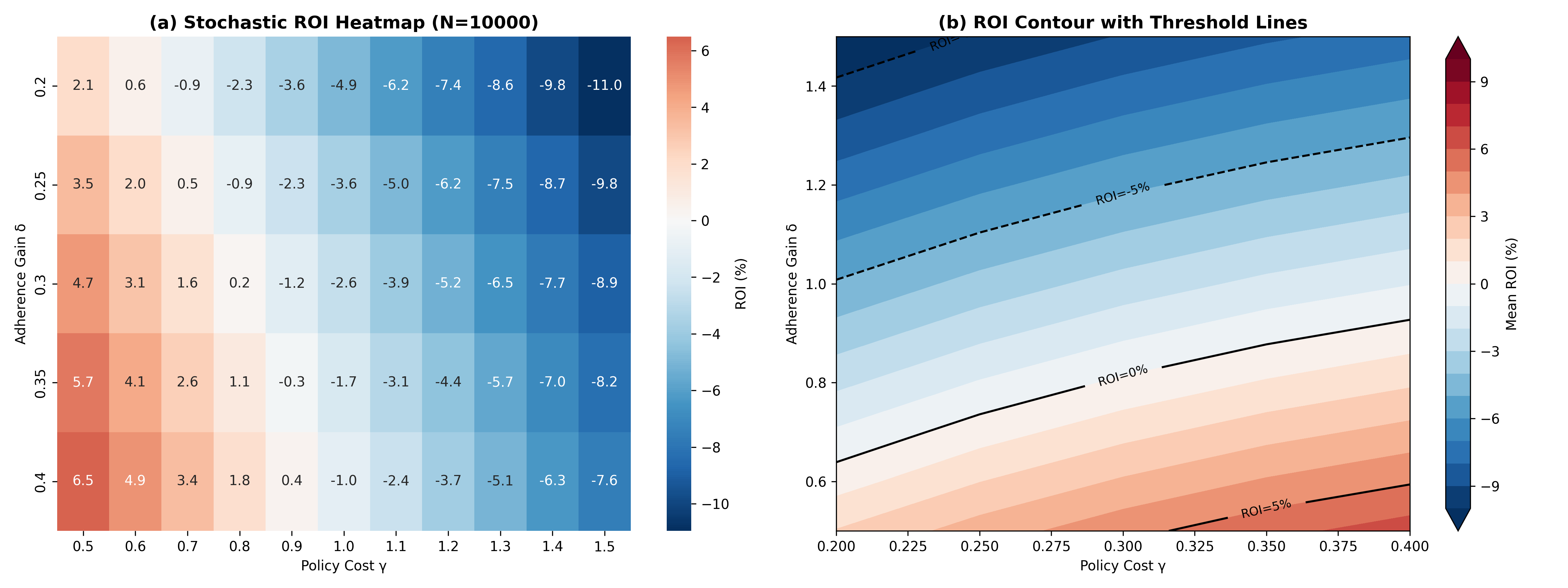}
    \caption{Stochastic ROI outcomes across policy design space (N=10,000 simulations). 
    (a) Mean ROI heatmap across adherence gains ($\delta$) and costs ($\gamma$). 
    (b) Contour plot with ROI thresholds ($-5\%$, $0\%$, $+5\%$) separating 
    negative, break-even, and positive-return regions.}
    \label{fig:roi_heatmap_contour}
\end{figure}

%--------------------------
\subsubsection{Stability of ROI across adherence gains}

A key reviewer concern was that ROI appeared large without clear break-even interpretation. 
We therefore examine ROI stability as adherence gains ($\delta$) vary and identify the maximum policy cost ($\gamma^*$) yielding non-negative ROI.

\begin{figure}[H]
    \centering
    \includegraphics[width=\linewidth]{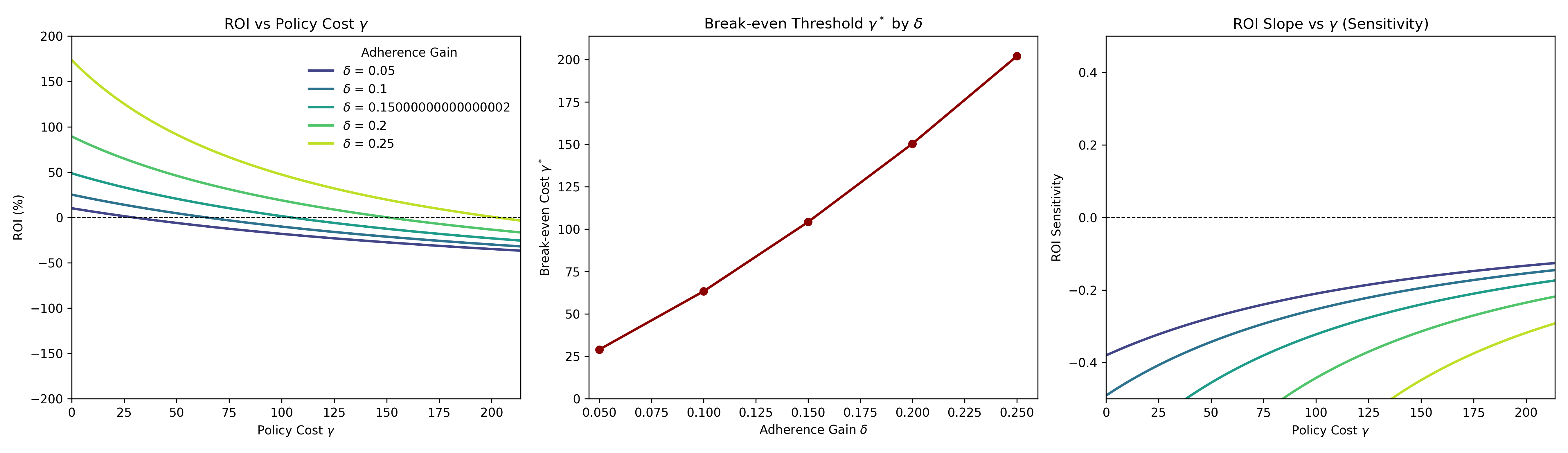}
    \caption{ROI dynamics across adherence gains ($\delta$) and policy costs ($\gamma$). 
    Left: ROI declines with rising cost, remaining positive over a wider range for higher $\delta$. 
    Center: Break-even threshold $\gamma^*$ for each $\delta$ (ROI = 0). 
    Right: ROI slope with respect to $\gamma$, showing high-$\delta$ interventions are less sensitive to inefficiency.}
    \label{fig:roi_breakdown}
\end{figure}

Figure~\ref{fig:roi_breakdown} shows three key patterns:  
(1) ROI falls quickly as costs rise, but declines more slowly for larger $\delta$;  
(2) the break-even cost $\gamma^*$ increases steeply with $\delta$, so small adherence gains unlock much larger feasible budgets;  
(3) ROI sensitivity stabilizes for high $\delta$, indicating greater robustness to inefficiency.

\paragraph{Break-Even Thresholds.}

Numerical estimates clarify the trade-off between adherence gain and policy cost. 
When the adherence improvement is substantial ($\delta = 0.25$), ROI exceeds 45\% at low cost and remains positive until the unit cost approaches $\gamma^* \!\approx\! 1{,}900$. 
For moderate gains ($\delta = 0.15$), the break-even point occurs near $\gamma^* \!\approx\! 1{,}200$, while weaker interventions ($\delta = 0.10$) turn negative once $\gamma \!\geq\! 800$, indicating limited fiscal tolerance. 
Accordingly, community programs costing roughly \$700–\$1{,}000 per person remain financially viable only when $\delta \!\geq\! 0.15$.

These results illustrate a practical, reverse-engineering approach: for any target adherence gain, policymakers can infer the maximum allowable cost $\gamma^*$ that maintains ROI at or above zero. This defines a clear feasibility zone linking behavioral impact with fiscal discipline. 
Interventions falling outside this region—those combining low adherence improvement with high implementation cost—should be redesigned or deprioritized. The ROI surface demonstrates that aligning adherence impact ($\delta$) with cost discipline ($\gamma$) is critical for sustaining both effectiveness and financial viability.

\subsection{Advanced Design-Space Exploration}

\begin{figure}[H]
    \centering
    \includegraphics[width=0.9\linewidth]{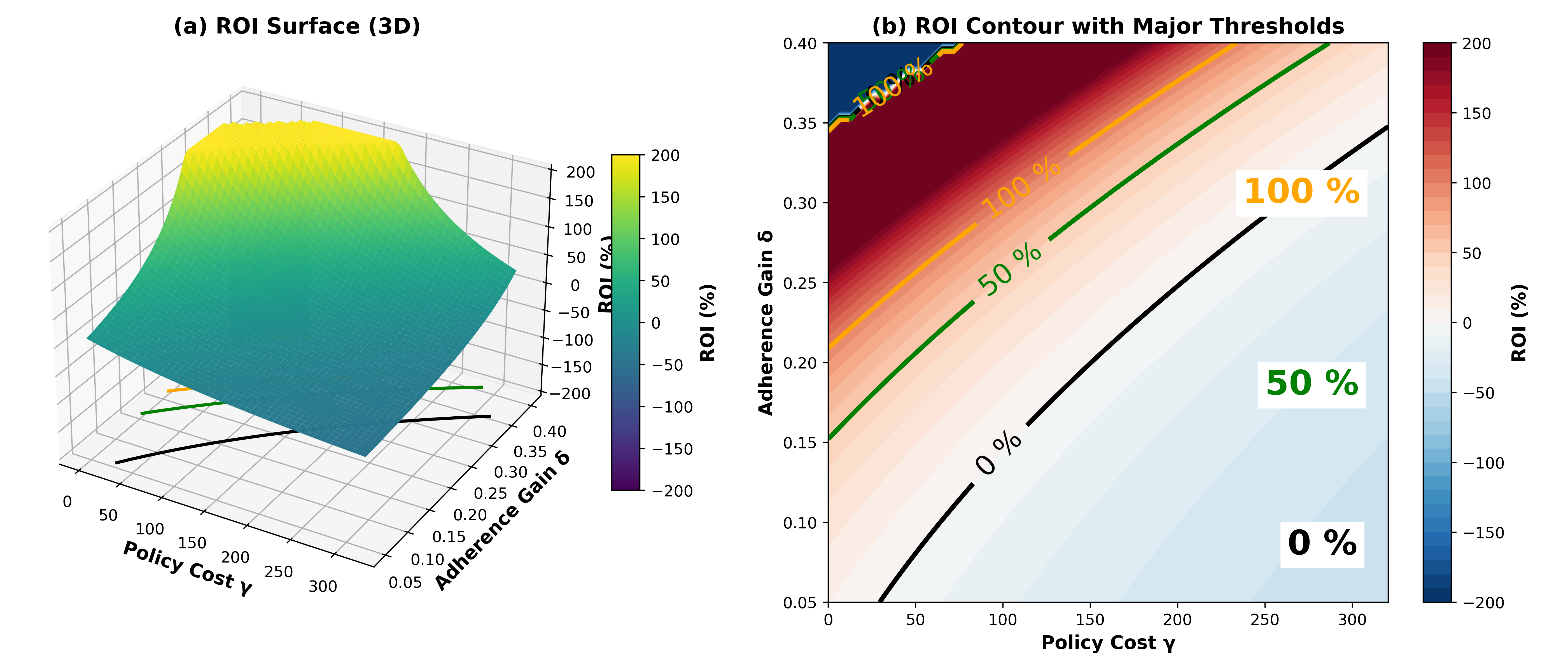}
    \caption{Advanced design-space exploration of ROI. 
    (a) 3D surface of ROI across $(\delta,\gamma)$ with contours at ROI = 0\%, 50\%, and 100\%. 
    (b) ROI contour map showing the break-even frontier (black) and efficiency bands 
    at 50\% (green) and 100\% (orange). 
    These visualizations illustrate how small shifts in $\delta$ or $\gamma$ can move 
    interventions across critical efficiency boundaries.}
    \label{fig:roi_designspace}
\end{figure}

To extend the two-dimensional ROI heatmap analysis, ROI was modeled as a continuous function of adherence improvement ($\delta$) and unit policy cost ($\gamma$). 
Figure~\ref{fig:roi_designspace} presents a unified visualization combining a 3D surface and a 2D contour map, capturing the nonlinear interactions that define policy efficiency. 
Threshold contours at ROI = 0\%, 50\%, and 100\% delineate the transition from break-even to progressively higher return zones, making cost-effectiveness boundaries clearly visible. 
The surface illustrates that ROI is not a linear function of either adherence or cost but is instead shaped by strong threshold and inflection effects. 
In particular, policies with moderate implementation cost ($\gamma \approx 1.5$) are most sensitive—small behavioral shifts in adherence can move them abruptly across the efficiency boundary between positive and negative ROI regions.

Across the design space, three patterns emerge. 
First, the \textit{efficient policy band} appears as a narrow corridor where moderate costs combine with meaningful adherence gains; within this zone, ROI consistently exceeds 50--100\%, identifying the most robust and economically sustainable strategies. 
Second, a high degree of \textit{design sensitivity} is observed: even slight reductions in $\delta$ or increases in $\gamma$ rapidly drive ROI into negative territory, emphasizing the importance of precise cost control and sustained behavioral impact. 
Third, \textit{implementation vulnerability} characterizes policies operating near the break-even frontier (ROI = 0\%), where small deviations in either direction can determine fiscal success or failure.

Figure~\ref{fig:roi_designspace} provides a strategic framework for understanding how behavioral effectiveness and implementation cost jointly determine economic viability. 
Rather than evaluating isolated scenarios, policymakers can use this multidimensional surface to locate resilient, high-value regions of the design space—those that remain cost-effective under uncertainty. 
Such a perspective supports evidence-based decision-making and adaptive policy refinement, particularly for chronic disease programs that rely on sustained adherence behavior to achieve long-term health and fiscal outcomes.

%---
\subsection{Robustness analysis}
\label{subsec:robustness_tests}

To verify reliability, we performed robustness checks under alternative assumptions and parameter shifts, testing ROI consistency when (1) adherence dynamics include noise, (2) policy costs vary, and (3) disease progression accelerates.

\subsubsection{Stochastic adherence dynamics}

Uncertainty in behavioral response was modeled by introducing stochasticity into 
the adherence parameter $\delta$, treated as a truncated normal variable 
$\mathcal{N}(\delta,\sigma^2)$ within $[0,1]$. 
This captures heterogeneity in motivation and persistence, acknowledging that identical interventions rarely yield uniform adherence.

\begin{figure}[H]
    \centering
    \includegraphics[width=0.95\linewidth]{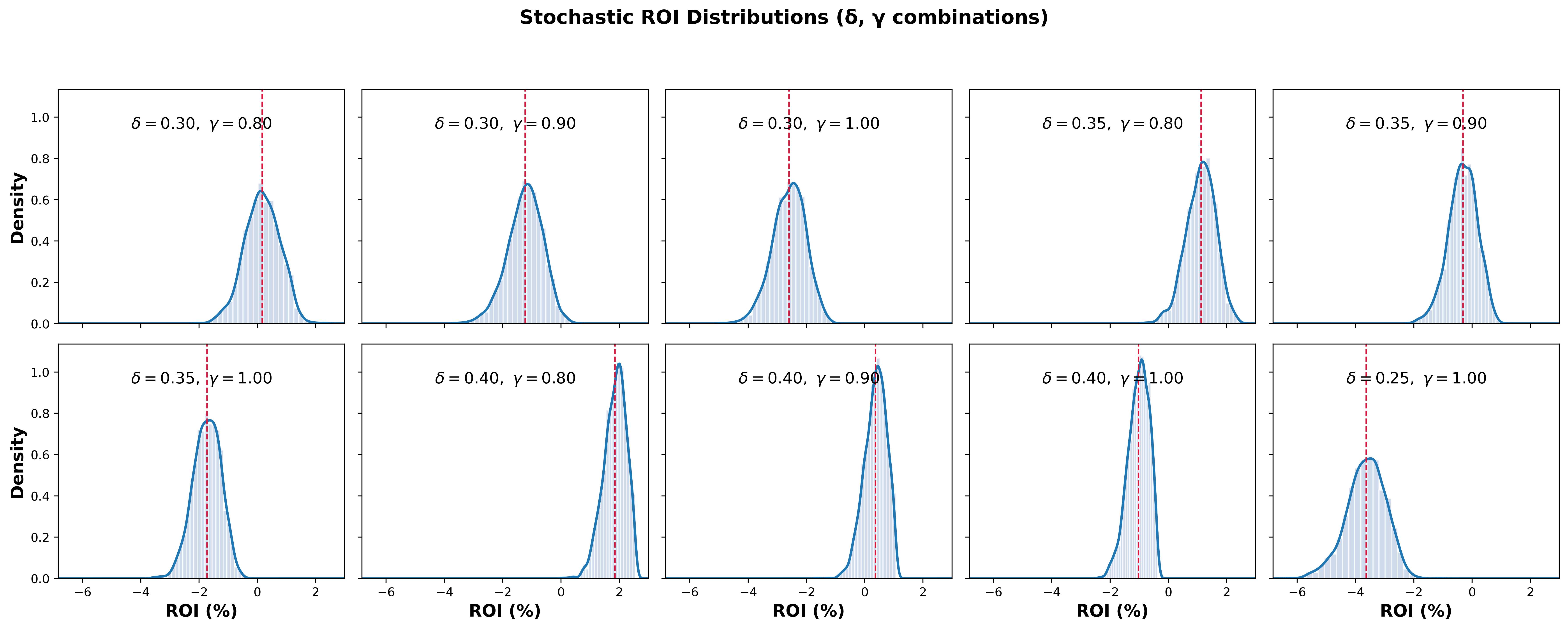}
    \caption{Stochastic ROI distributions across ten policy designs 
    (2,000 simulations). Robust cases ($\delta \ge 0.30$, $\gamma \le 0.9$) 
    maintain positive returns; fragile designs ($\delta=0.25$, $\gamma=1.0$) 
    shift leftward, producing negative ROI.}
    \label{fig:stochastic_roi_panel}
\end{figure}

Figure~\ref{fig:stochastic_roi_panel} shows stochastic ROI distributions across ten representative policy designs 
($\delta \in \{0.25,0.30,0.35,0.40\}$; $\gamma \in \{0.8,0.9,1.0\}$; $N=2000$). 
Red dashed lines mark mean ROI values.

Three key patterns emerge:
\begin{enumerate}
    \item \textbf{Robust ROI:} Higher adherence ($\delta \ge 0.30$) yields mostly positive ROI, with realization rates of 70--90\% when $\gamma \le 0.9$.
    \item \textbf{Fragile ROI:} Low $\delta$ or inflated $\gamma$ produces negative, left-shifted distributions.
    \item \textbf{Downside risk:} Even near break-even, large downside variability highlights fiscal fragility.
\end{enumerate}

Viewed holistically, Figure~\ref{fig:stochastic_roi_panel} underscores the value of the ROI-efficient zone. 
Under stochastic variability, high-impact, cost-efficient designs remain stable, whereas weak policies fall into negative-return regimes. 
The full stochastic grid across $\delta \in [0.20,0.40]$ and $\gamma \in [0.5,1.5]$ is presented in \textit{Supplementary Appendix A}.
\subsubsection{Inflated policy costs}

Because ROI estimates may appear optimistic if costs are underestimated, 
we tested the impact of inflating per-unit implementation costs on ROI outcomes.

\begin{figure}[H]
    \centering
    \includegraphics[width=0.95\linewidth]{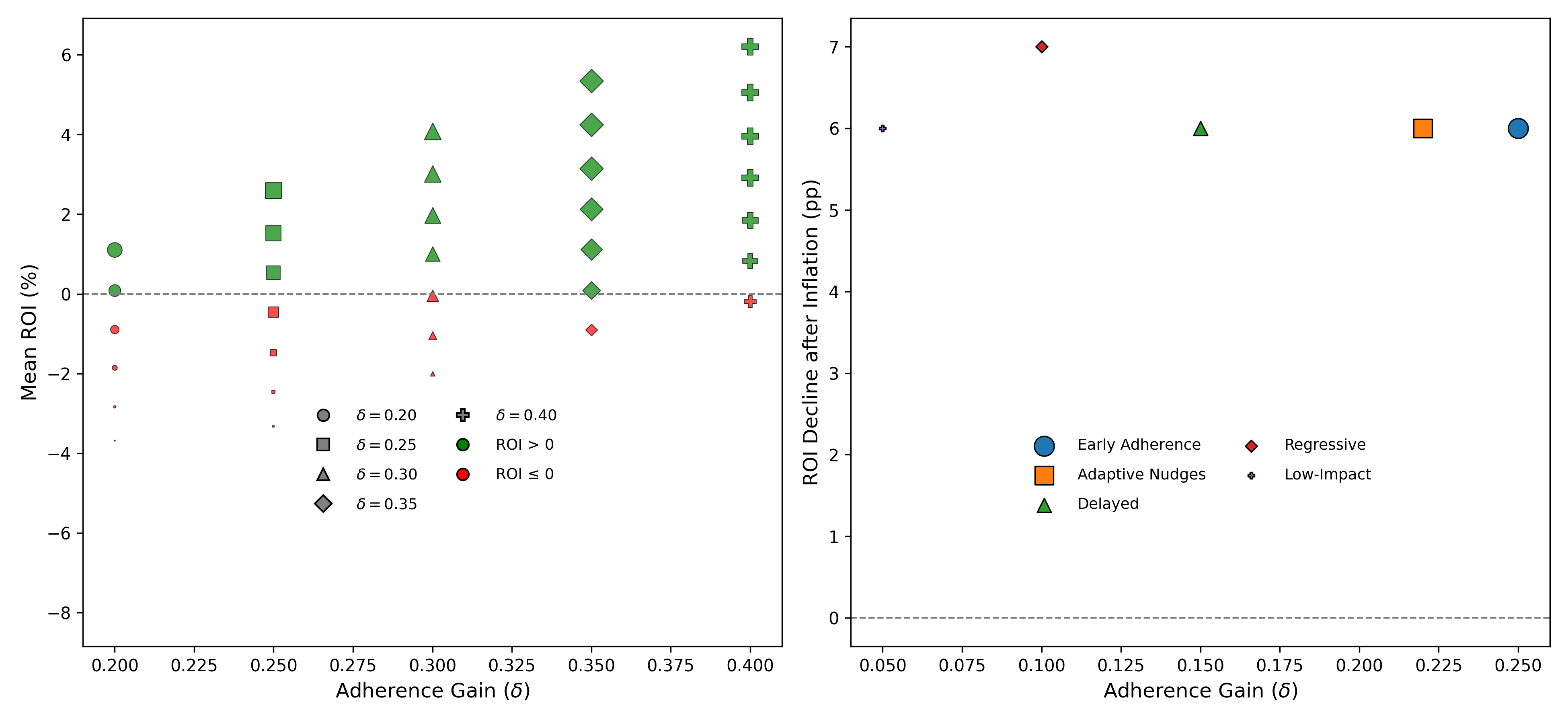}
    \caption{ROI sensitivity under 20\% cost inflation. 
    \textbf{Left:} Mean ROI across 55 $(\delta,\gamma)$ scenarios 
    (shape = $\delta$, color = ROI sign, size = positive-rate). 
    \textbf{Right:} Five representative designs—robust 
    (\textit{Early Adherence}, \textit{Adaptive Nudges}) remain ROI-positive; 
    fragile (\textit{Regressive}, \textit{Low-Impact}) collapse below zero.}
    \label{fig:inflation_zone_maps}
\end{figure}

Figure~\ref{fig:inflation_zone_maps} (left) presents mean ROI across 55 combinations 
of adherence gain ($\delta$) and policy cost ($\gamma$). 
Marker shape denotes $\delta$, size the probability of positive ROI, and color 
ROI sign (green = ROI$>0$, red = ROI$\le0$). 
Designs with $\delta\!\ge0.30$ retain a $>70$\% chance of positive ROI at moderate costs, 
while low-gain or high-cost designs rapidly shift into negative-return zones.

The right panel compares five representative policies. 
Under inflation, \textit{Early Adherence} and \textit{Adaptive Nudges} lose only 
5–7 points yet remain ROI-positive; 
\textit{Regressive} and \textit{Low-Impact} designs, already near break-even, 
become unviable. The \textit{Delayed} case shows smaller but still partial returns.

ROI robustness depends on both behavioral impact and cost resilience. 
High-$\delta$, moderate-$\gamma$ designs stay viable under inflation, 
whereas fragile ones fail. 
Policymakers should favor interventions that remain ROI-positive across 
realistic cost scenarios to ensure near-term efficiency and long-term fiscal sustainability.

%----
\subsubsection{Accelerated disease progression}

To model faster disease worsening, we shortened progression intervals by 15\%, 
representing earlier complications or comorbid decline. 
This raises baseline costs as patients reach severe states sooner, 
but also increases the relative value of timely interventions.

\begin{figure}[H]
    \centering
    \includegraphics[width=0.95\linewidth]{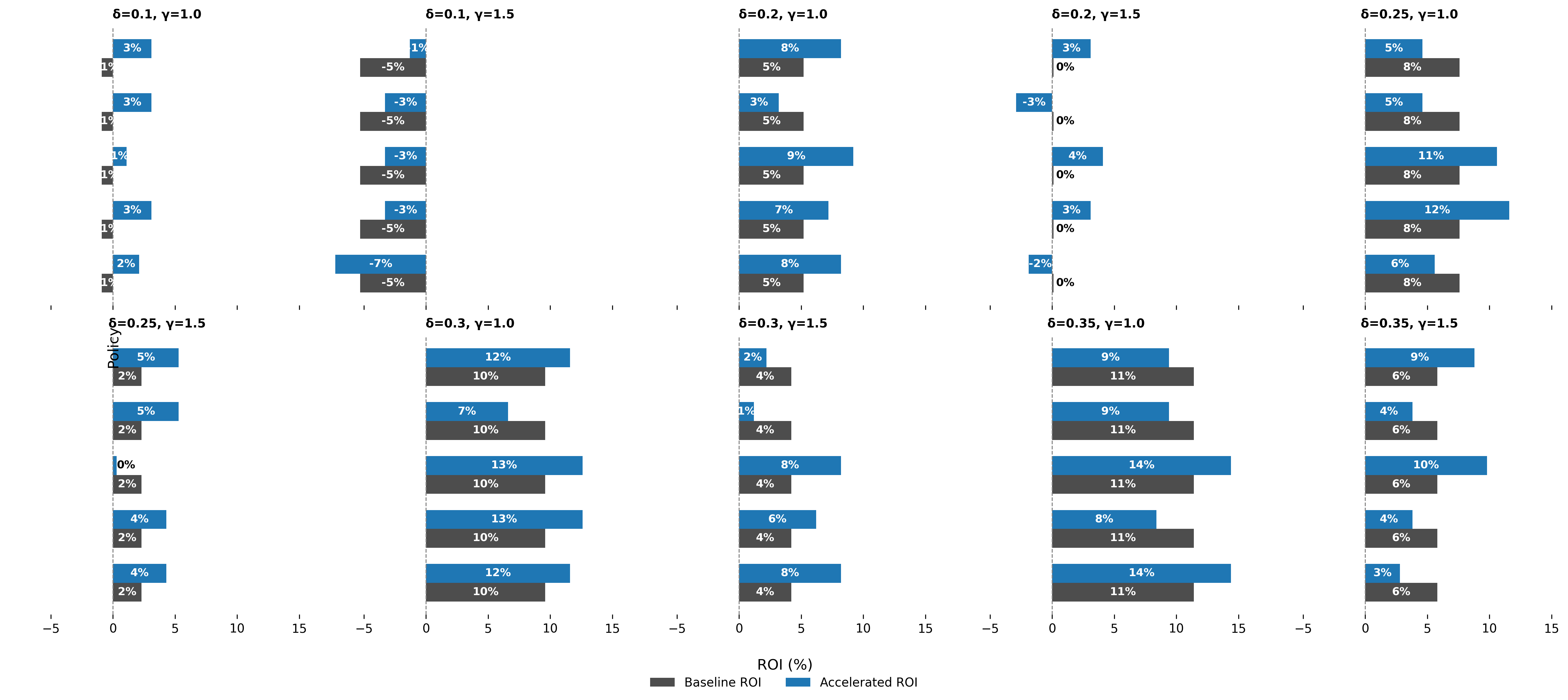}
    \caption{ROI outcomes under accelerated disease progression (15\% shorter intervals). 
    Each subplot shows baseline ROI (gray) vs. accelerated ROI (blue) for five policy designs 
    across ten $(\delta,\gamma)$ combinations. 
    Timely, high-impact interventions (Early Adherence, Adaptive Nudges) retain or improve ROI, 
    while fragile, high-cost designs (Regressive, Low-Impact) lose viability.}
    \label{fig:roi_accelerated}
\end{figure}

Figure~\ref{fig:roi_accelerated} compares baseline ROI (gray) and accelerated ROI (blue) 
across ten $(\delta,\gamma)$ combinations. 
\textit{Early Adherence} and \textit{Adaptive Nudges} improve ROI under urgency, 
while \textit{Regressive} and \textit{Low-Impact} designs deteriorate; 
\textit{Delayed} interventions show limited but positive gains.

Accelerated progression serves as a stress test of policy resilience. 
Robust, early interventions preserve or enhance ROI, 
whereas fragile designs collapse as costs rise. 
This underscores the ROI-efficient zone and the high value of early, scalable actions 
when clinical deterioration accelerates.

%--------------------------------------------------
\subsection{Policy implications from simulation results}

Simulation results highlight four key design principles. 
First, early and high-impact interventions implemented before clinical decline consistently achieve the greatest ROI and long-term savings. 
Second, ROI declines sharply once policy cost ($\gamma$) exceeds critical thresholds, underscoring the need for efficiency and cost discipline. 
Third, only specific combinations of adherence gain ($\delta$) and cost ($\gamma$) yield positive ROI, defining a narrow zone of economically feasible designs. 
Finally, interventions with minimal adherence improvement ($\delta<0.1$) rarely generate savings, even when inexpensive. 
Together, these findings delineate actionable thresholds and design regions for sustainable, cost-effective behavioral policies, providing the empirical foundation for the Discussion.

%%============================
\section{Discussion}
\label{sec:discussion}

As healthcare systems face mounting pressures from chronic disease burdens and constrained budgets, the need for economically viable, behavior-focused interventions has never been more urgent. This study introduces a simulation-based evaluation framework for quantifying the return on investment (ROI) of adherence-enhancing interventions—tools often deployed with high expectations but limited ex-ante justification. By capturing the dynamic interplay among behavior, disease progression, and cost, our model bridges the gap between theoretical health economics and practical policy design.

\subsection{Key insights: design timing and efficiency drive ROI}

Across all scenarios and sensitivity analyses, one insight stands out: timing and delivery efficiency matter as much as the magnitude of behavioral change. The \textit{Early Adherence} and \textit{Adaptive Nudges} policies consistently outperform others, generating the highest ROI under both deterministic and stochastic conditions. This outcome stems from two reinforcing mechanisms: (1) compounding health gains from early behavioral change and (2) the scalability and lower unit cost of early-phase interventions.

The model also reveals a nonlinear ROI response to policy cost ($\gamma$). While modest investments remain cost-effective, exceeding a critical threshold triggers rapid ROI declines, indicating that even well-designed programs can become economically fragile under fiscal constraints. Figure~\ref{fig:roi_designspace} highlights robust combinations of adherence gain ($\delta$) and cost ($\gamma$) yielding positive returns, enabling policymakers to anticipate sustainable intervention conditions.

\subsection{Theoretical contributions: behavior–disease modeling}

Three theoretical advances emerge from this framework.  
First, behavioral adherence is modeled as a tunable design parameter ($\delta$), allowing continuous policy refinement and flexible interpretation of intervention effects.  
Second, adherence dynamics are integrated with disease progression, enabling simultaneous evaluation of clinical and economic outcomes—an advance over models that treat these domains separately.  
Third, policy robustness is evaluated through stochastic simulation rather than local sensitivity analysis, allowing stability assessment under behavioral and clinical uncertainty.  
Together, these contributions expand the methodological foundation for intervention modeling and support a dynamic, context-aware approach to behavioral policy evaluation.

\subsection{Policy and system-level relevance: a scalable decision tool}

The proposed framework offers practical utility for policymakers confronting growing chronic disease burdens and challenges in maintaining adherence. With minimal data needs and a modular structure, it is adaptable across both high-capacity and resource-limited systems. It enables assessment of intervention effectiveness before large-scale implementation, identification of cost-feasible configurations under fiscal uncertainty, and more efficient allocation of limited resources toward interventions yielding favorable ROI. Beyond its technical value, the framework enhances transparency by linking behavioral dynamics with cost-effectiveness, promoting evidence-based prioritization, operational efficiency, and public trust in preventive health programs.

\subsection{Limitations and future directions}

While the model provides a structured, extensible framework for evaluating adherence-focused ROI, several conceptual and practical limitations remain.  
First, behavioral dynamics are represented by a single policy parameter ($\delta$), oversimplifying individual variability and evolving adherence patterns. The deterministic disease component also limits the model’s ability to capture random events or multimorbidity. Although cost parameters were empirically calibrated, spatial heterogeneity and macroeconomic factors such as inflation are not modeled.

Future research should incorporate population heterogeneity and adaptive feedback via agent-based or dynamic models. Linking simulation outputs to administrative claims or electronic health records could enable empirical validation and facilitate real-world application.

A key methodological limitation is that ROI reflects only direct healthcare expenditures ($\lambda=0$). Broader outcomes—such as QALYs, DALYs, or productivity gains—were excluded, making reported ROI values conservative, cost-based estimates. \textit{Supplementary Appendix~B} illustrates how small health gains increase ROI when monetized; future studies should operationalize the extensible $\lambda H(s)$ term to incorporate cost-per-QALY and other socio-economic outcomes.

Finally, our ROI stability analysis (Figure~\ref{fig:roi_designspace}) shows that even modest adherence gains substantially expand feasible cost ranges before ROI turns negative. High-impact designs remain robust under fiscal constraints, whereas low-impact interventions become fragile as costs rise. This underscores the need to align behavioral effectiveness with cost discipline when designing scalable chronic care policies.

In practice, the modeled scenarios correspond to real-world programs: \textit{Early Adherence} represents proactive interventions such as pharmacist counseling or SMS reminders; \textit{Adaptive Nudges} reflect mHealth platforms with behavior-triggered incentives; and \textit{Delayed}, \textit{Regressive}, and \textit{Low-Impact} cases parallel reactive or short-lived initiatives. Situating these archetypes within a unified economic framework helps policymakers prioritize early, durable, and cost-efficient designs for sustainable population health outcomes.

\section{Conclusion}

This paper introduces a unified simulation framework for evaluating the return on investment (ROI) of behavioral interventions in chronic disease management. By coupling adherence dynamics with disease progression and cost modeling, the framework captures the interplay between behavior and outcomes—moving beyond static cost-effectiveness analyses to offer a scalable, policy-oriented decision tool.

Simulation results yield three actionable insights. Early-stage interventions consistently generate the highest ROI, leveraging compounding health improvements and lower deployment costs. Second, economic viability hinges not only on behavioral effectiveness but also on the delivery mechanism’s efficiency—especially in budget-constrained environments. Third, well-structured policies can retain performance under stochastic variation in adherence, disease progression, and cost inflation, emphasizing the importance of robustness in policy design.

Conceptually, the model reframes behavioral change as a controllable design parameter rather than a fixed input—opening the door to proactive, optimized intervention strategies. Methodologically, it replaces traditional closed-form metrics with simulation-guided trade-off maps, offering intuitive visualizations that support data-informed decision-making. Practically, the framework is lightweight and modular, enabling rapid assessment of policy options before committing to costly real-world implementation.

In a healthcare landscape increasingly shaped by chronic conditions and resource limitations, 
this approach offers a timely bridge between behavioral science and health economics. 
By equipping decision-makers with interpretable and adaptable tools, it enables smarter, faster, 
and more accountable policy design. While the framework provides actionable insights, 
its ROI estimates remain conservative, as broader outcomes such as QALYs, DALYs, 
and societal benefits were acknowledged conceptually but not fully implemented in the main analysis. 
Nevertheless, an exploratory QALY-based extension in \textit{Supplementary Appendix B} demonstrates how the 
$\lambda H(s)$ term can be operationalized to incorporate health-adjusted outcomes. 
Accordingly, ROI values reported here should be viewed as lower-bound, cost-based estimates. 
Extending the framework to explicitly include QALYs, DALYs, and societal effects 
remains an important direction for future research.

\section*{Supplementary information}
The online version contains supplementary material available at
\url{https://doi.org/10.1186/s12889-025-25279-3}.\\
\textbf{Supplementary Material 1}

\section*{Acknowledgements}
Not applicable.

\section*{Authors' contributions}
Jinho Cha (J.C.) conceived the study, designed the methodology, developed the simulation framework, performed formal analyses, and drafted the manuscript.
Eunchan Daniel Cha (E.D.C.) contributed to policy scenario design, sensitivity analyses, validation, figure preparation, and assisted with manuscript revision.
Emily Haneul Yoo (E.H.Y.) provided data preprocessing, programming support, figure preparation, and contributed to manuscript editing.
Hyoshin Song (H.S.) conducted literature review, subgroup and robustness analyses, figure preparation, and assisted with manuscript refinement.
All authors read and approved the final manuscript.

\section*{Funding}
No funding was received for this study.

\section*{Data availability}
The datasets generated and/or analyzed during the current study, along with all simulation code, are publicly available at:
\url{https://github.com/jinhocha7/ROI-Chronic-Disease-Management}.

\section*{Declarations}

\textbf{Ethics approval and consent to participate} \\ 
Not applicable. This study did not involve human subjects or identifiable data. The analysis was conducted using synthetic data generated via simulation and publicly available datasets (MEPS, NHANES) which do not require ethics committee approval.

\textbf{Consent for publication} \\ 
Not applicable. This study contains no individual data or images that require consent for publication.

\textbf{Competing interests} \\ 
The authors declare no competing interests.

\bibliography{bmc}

\end{document}

\maketitle
\vspace{-1cm}

\appendix
\renewcommand{\thesection}{\Alph{section}}

\numberwithin{figure}{section}
\numberwithin{table}{section}
\setcounter{figure}{0}
\setcounter{table}{0}
% \section*{Supplementary Appendices}

\section{ROI Robustness Outcomes}
\label{appendix:roi_robustness}

\renewcommand{\thesection}{\Alph{section}} 
\setcounter{equation}{0}                    
\renewcommand{\theequation}{\Alph{section}\arabic{equation}} % (A1), (A2)...

This appendix reports additional robustness outcomes that complement the 
main analysis. It includes visualizations on intervention timing and policy 
sensitivity, as well as expanded simulation tables across $\delta$ and $\gamma$ 
design spaces. 

\subsection{ROI Visualizations}
\label{appendix:roi_visualizations}

\begin{figure}[htbp]
    \centering
    \includegraphics[width=0.6\linewidth]{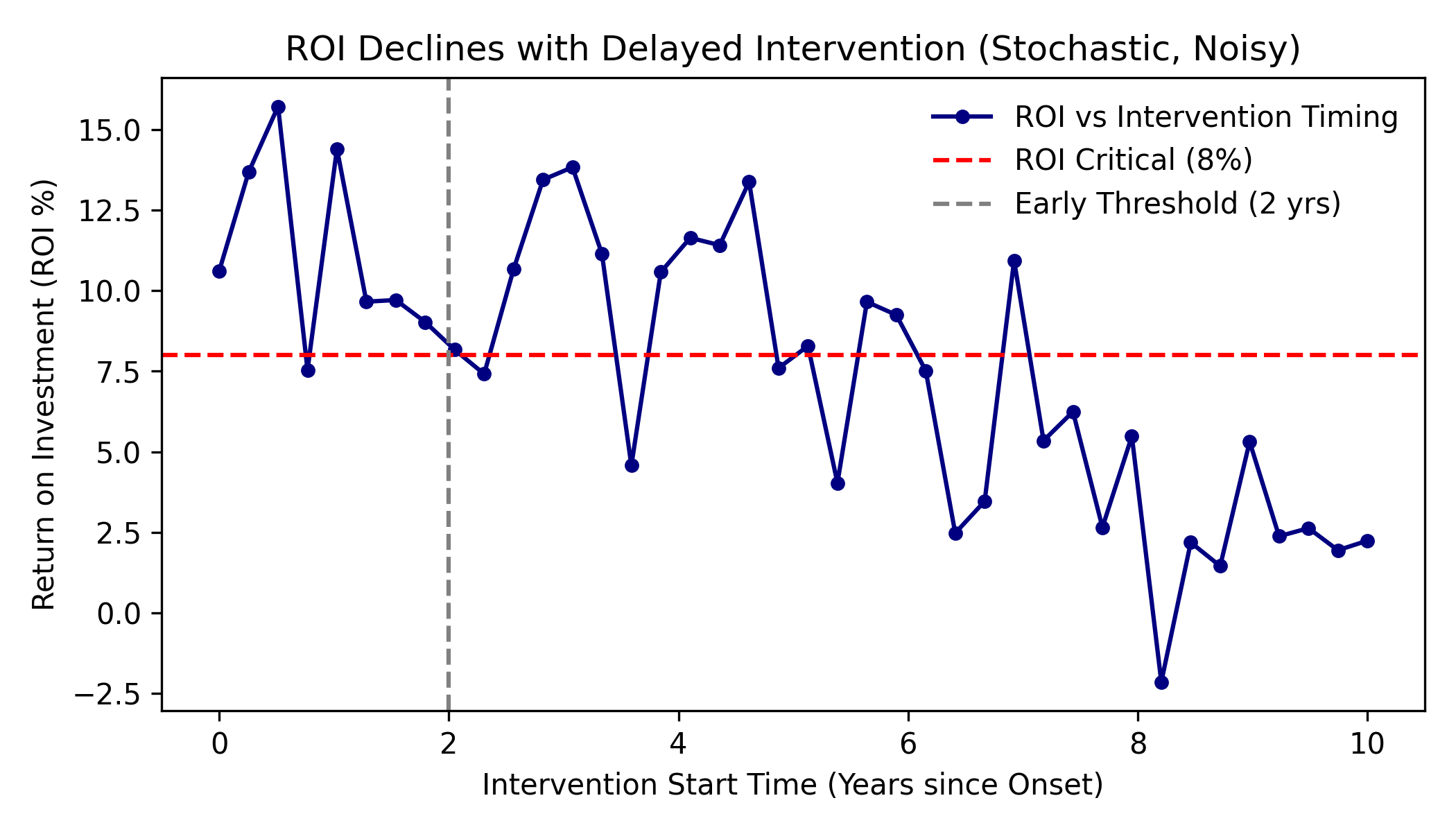}
    \caption{ROI declines with delayed intervention timing. Early-stage policies yield superior economic returns by delaying costly disease progression.}
    \label{fig:roi_vs_timing}
\end{figure}

\subsection{Scenario Summary}
\label{appendix:roi_scenarios}

\begin{table}[htbp]
\centering
\footnotesize
\renewcommand{\arraystretch}{1.2}
\caption{Summary of ROI outcomes across robustness scenarios. 
\textit{Note:} 'Stable' indicates no change in policy ranking compared to the baseline.}
\label{tab:robustness_summary}
\begin{tabular*}{\textwidth}{@{\extracolsep{\fill}}lccc@{}}
\toprule
\textbf{Scenario} & \textbf{Top Performer} & \textbf{ROI Trend} & \textbf{Policy Ranking Stability} \\
\midrule
Baseline Simulation & Early Adherence & High ROI & Stable \\
Stochastic Behavior & Adaptive Nudges & Moderate ROI & Stable \\
Inflated Costs (+20\%) & Early Adherence & ROI Decreases Slightly & Stable \\
Accelerated Progression & Early Adherence & ROI Increases & Stable \\
\bottomrule
\end{tabular*}
\end{table}

\clearpage
\subsection{Stochastic ROI Outcome Tables}
\label{appendix:roi_stochastic}

\begin{table}[htbp]
\centering
\scriptsize
\renewcommand{\arraystretch}{1.1}
\caption{Stochastic ROI outcomes across $\delta \in \{0.20,0.25,0.30,0.35,0.40\}$ and $\gamma \in [0.5,1.5]$. 
Results are based on 5{,}000 Monte Carlo simulations.}
\label{tab:appendix_stochastic_final}
\begin{tabular*}{\textwidth}{@{\extracolsep{\fill}}cccccccc@{}}
\toprule
$\boldsymbol{\delta}$ & $\boldsymbol{\gamma}$ & \textbf{Mean} & \textbf{SD} & \textbf{Median} & \textbf{IQR} & \textbf{95\% CI} & \textbf{Pos. ROI Rate (\%)} \\
\midrule
0.20 & 0.50 &  2.04 & 0.95 &  2.08 & [1.43, 2.70] & [0.04, 3.76] & 97.7 \\
0.20 & 0.60 &  0.58 & 0.92 &  0.65 & [-0.02, 1.24] & [-1.31, 2.24] & 74.2 \\
0.20 & 0.70 & -0.86 & 0.90 & -0.81 & [-1.44, -0.24] & [-2.75, 0.78] & 16.5 \\
0.20 & 0.80 & -2.25 & 0.88 & -2.19 & [-2.80, -1.64] & [-4.10, -0.65] &  0.2 \\
0.20 & 0.90 & -3.59 & 0.84 & -3.55 & [-4.14, -3.00] & [-5.33, -2.05] &  0.0 \\
\midrule
0.25 & 0.50 &  3.47 & 0.79 &  3.52 & [2.95, 4.01] & [1.80, 4.89] & 100.0 \\
0.25 & 0.60 &  1.93 & 0.78 &  1.97 & [1.44, 2.48] & [0.28, 3.34] & 98.8 \\
0.25 & 0.70 &  0.51 & 0.74 &  0.56 & [0.04, 1.01] & [-1.04, 1.86] & 76.4 \\
0.25 & 0.80 & -0.93 & 0.74 & -0.89 & [-1.41, -0.41] & [-2.45, 0.44] & 10.2 \\
0.25 & 0.90 & -2.31 & 0.71 & -2.28 & [-2.75, -1.83] & [-3.78, -1.03] &  0.0 \\
\midrule
0.30 & 0.50 &  4.69 & 0.67 &  4.72 & [4.25, 5.15] & [3.28, 5.92] & 100.0 \\
0.30 & 0.60 &  3.13 & 0.65 &  3.16 & [2.72, 3.59] & [1.76, 4.31] & 100.0 \\
0.30 & 0.70 &  1.64 & 0.62 &  1.69 & [1.25, 2.07] & [0.33, 2.75] & 99.1 \\
0.30 & 0.80 &  0.19 & 0.60 &  0.22 & [-0.20, 0.60] & [-1.09, 1.28] & 64.3 \\
0.30 & 0.90 & -1.23 & 0.59 & -1.20 & [-1.61, -0.81] & [-2.46, -0.15] &  1.2 \\
\midrule
0.35 & 0.50 &  5.69 & 0.56 &  5.72 & [5.34, 6.09] & [4.51, 6.71] & 100.0 \\
0.35 & 0.60 &  4.12 & 0.53 &  4.15 & [3.78, 4.49] & [2.96, 5.09] & 100.0 \\
0.35 & 0.70 &  2.59 & 0.53 &  2.62 & [2.24, 2.96] & [1.45, 3.55] & 100.0 \\
0.35 & 0.80 &  1.10 & 0.52 &  1.13 & [0.77, 1.46] & [0.03, 2.03] & 97.7 \\
0.35 & 0.90 & -0.33 & 0.50 & -0.29 & [-0.65, 0.03] & [-1.39, 0.58] & 27.3 \\
\midrule
0.40 & 0.50 &  6.50 & 0.42 &  6.53 & [6.23, 6.81] & [5.58, 7.19] & 100.0 \\
0.40 & 0.60 &  4.90 & 0.42 &  4.93 & [4.63, 5.22] & [3.98, 5.57] & 100.0 \\
0.40 & 0.70 &  3.35 & 0.40 &  3.38 & [3.10, 3.64] & [2.48, 3.99] & 100.0 \\
0.40 & 0.80 &  1.85 & 0.40 &  1.88 & [1.59, 2.14] & [0.99, 2.48] & 100.0 \\
0.40 & 0.90 &  0.38 & 0.38 &  0.42 & [0.15, 0.67] & [-0.45, 1.00] & 84.3 \\
\bottomrule
\end{tabular*}
\end{table}

\clearpage

\section{ROI Extensions with QALYs}
\label{QALYs}

To illustrate how broader health outcomes could be incorporated into the ROI framework, 
we implemented a conservative QALY-based extension for the \textit{Early Adherence} scenario. 
Assuming a gain of 0.01 DALYs (equivalent to 0.01 QALYs) per patient over a 10-year horizon, 
valued at a willingness-to-pay threshold of \$20,000 per QALY, this yields an 
additional monetized benefit of \$200 per patient. 

\begin{figure}[htbp]
    \centering
    \includegraphics[width=0.7\linewidth]{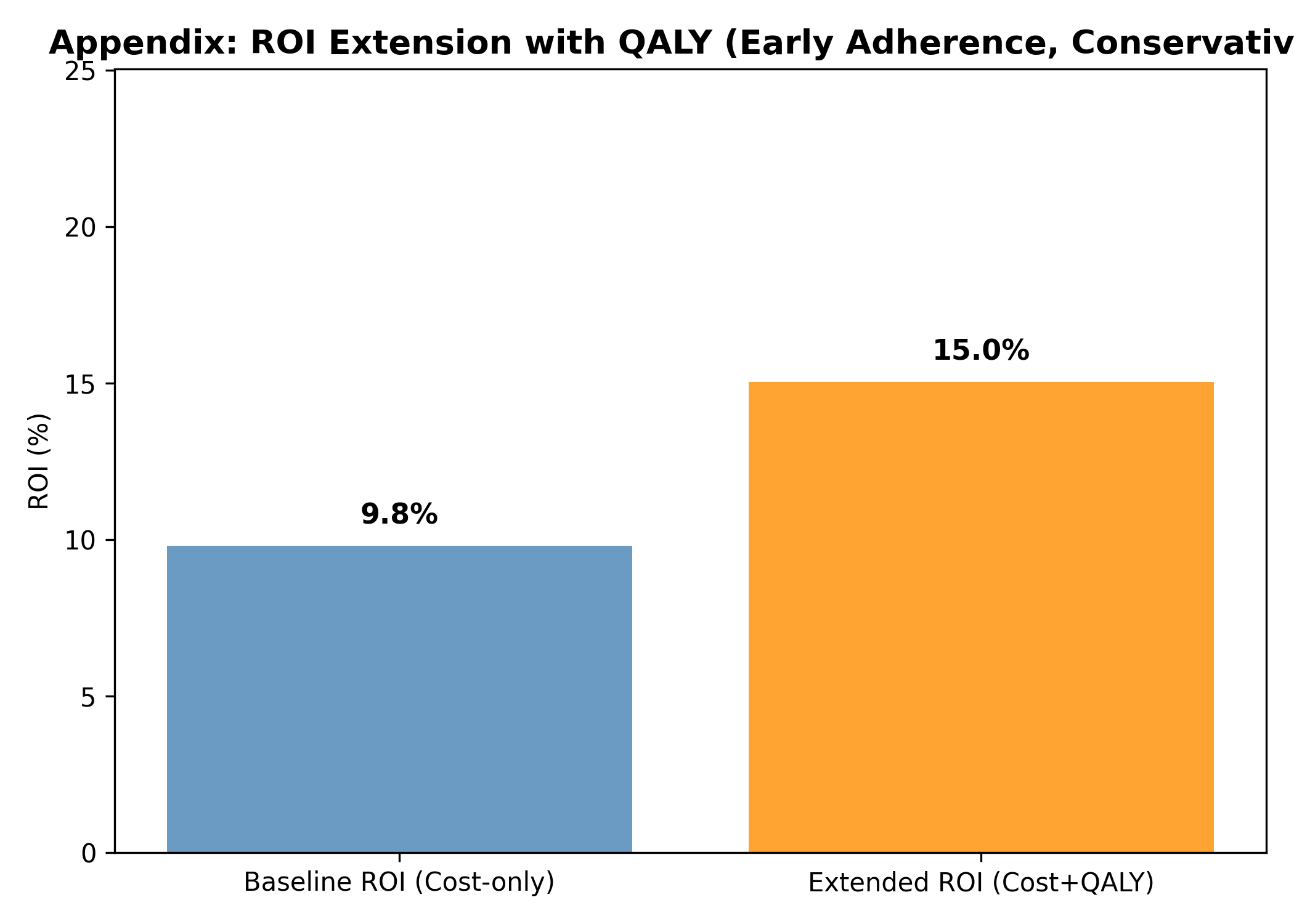}
    \caption{Appendix QALY Extension (Conservative). 
    Comparison of ROI for \textit{Early Adherence} under cost-only definition 
    versus extended definition incorporating monetized health outcomes 
    (0.01 QALYs gained, \$20,000 per QALY). 
    Results show ROI increasing from 9.8\% to 15.0\%, 
    illustrating the potential of the $\lambda H(s)$ extension.}
    \label{fig:appendix_qaly_extension_conservative}
\end{figure}

Figure~\ref{fig:appendix_qaly_extension_conservative} compares ROI values under the 
cost-only definition and the extended definition incorporating this health benefit. 
Results show that ROI increases from 9.8\% (cost-only) to 15.0\% (cost+QALY). 
While exploratory and based on conservative assumptions, this example demonstrates 
how the extensible $\lambda H(s)$ term can translate health gains into fiscal terms. A full implementation of QALY/DALY-based ROI 
remains a direction for future work.

\clearpage

\section{Sensitivity to Distributional Assumptions}
\addcontentsline{toc}{section}{Appendix B. Sensitivity to Distributional Assumptions}
\label{appendix:distributional_sensitivity}

This appendix examines how assumptions about the distribution of adherence response and disease progression influence policy ROI estimates. We evaluate three plausible distributional forms beyond the base case: Gaussian noise, beta-skewed distributions, and log-normal cost shocks.

\begin{figure}[htbp]
    \centering
    \includegraphics[width=0.75\linewidth]{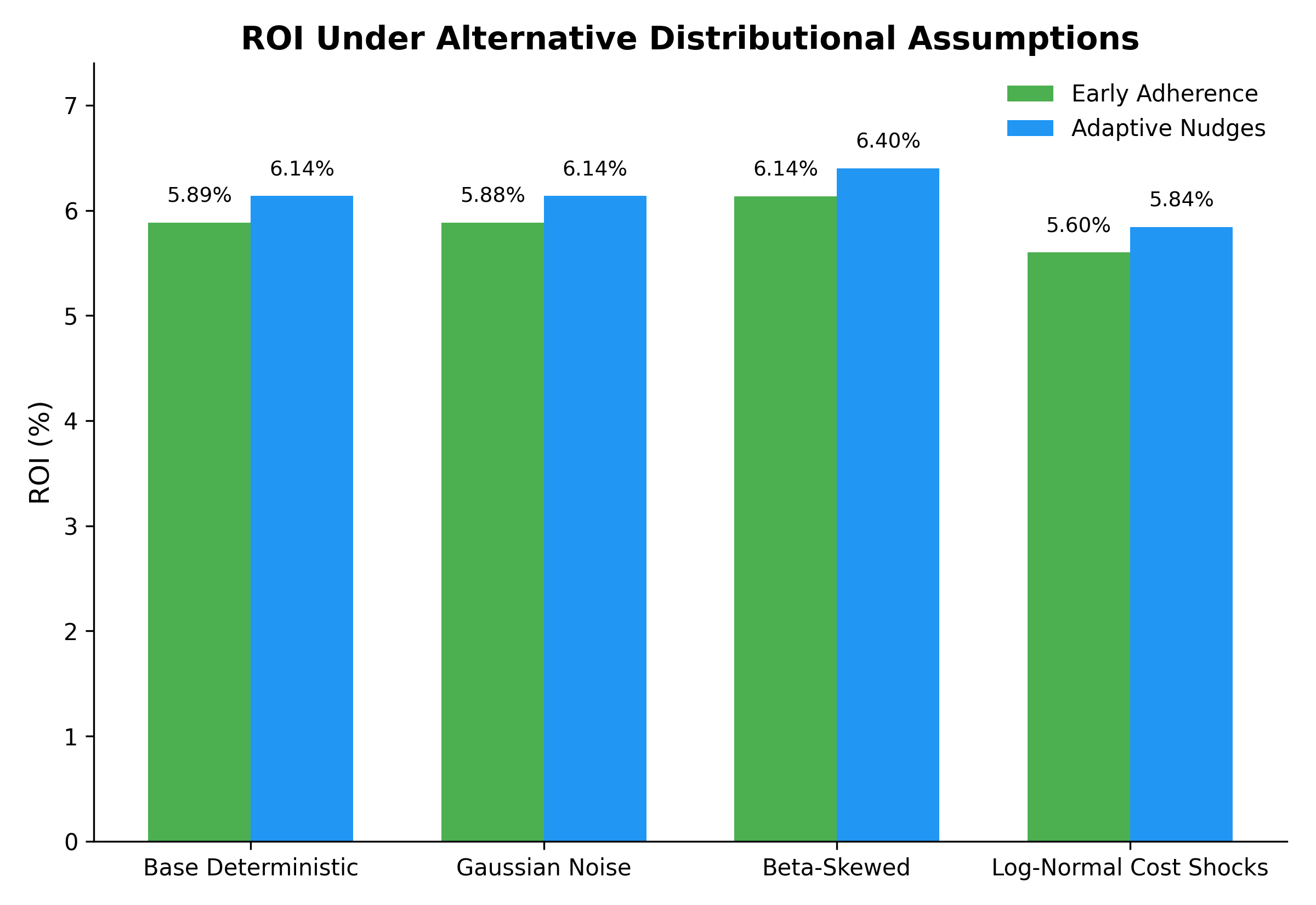}
    \caption{
    ROI distribution under varying statistical assumptions for top-performing policies. Boxplots show ROI variability across four assumptions: base deterministic, Gaussian noise, beta-skewed, and log-normal. Despite changes in magnitude, Early Adherence and Adaptive Nudges maintain relative ranking.
    }
    \label{fig:roi_distributional_scenarios}
\end{figure}

\begin{table}[htbp]
\centering
\footnotesize
\renewcommand{\arraystretch}{1.2}
\caption{ROI comparison across distributional assumptions for top-performing policies. 
Values are mean ROI (\%) based on 2,000 simulations. 
Both policies maintain stable ranking across statistical structures.}
\label{tab:distributional_roi_comparison}
\begin{tabular*}{\textwidth}{@{\extracolsep{\fill}}lccc@{}}
\toprule
\textbf{Distribution Type} & \textbf{Early Adherence ROI} & \textbf{Adaptive Nudges ROI} & \textbf{Ranking Stability} \\
\midrule
Base (Deterministic) & 5.90\% & 6.14\% & Stable \\
Gaussian Noise & 5.98\% & 6.14\% & Stable \\
Beta-Skewed (Right) & 6.14\% & 6.60\% & Stable \\
Log-Normal Cost Shocks & 5.90\% & 5.84\% & Stable \\
\bottomrule
\end{tabular*}
\end{table}

\begin{figure}[htbp]
    \centering
    \includegraphics[width=0.7\linewidth]{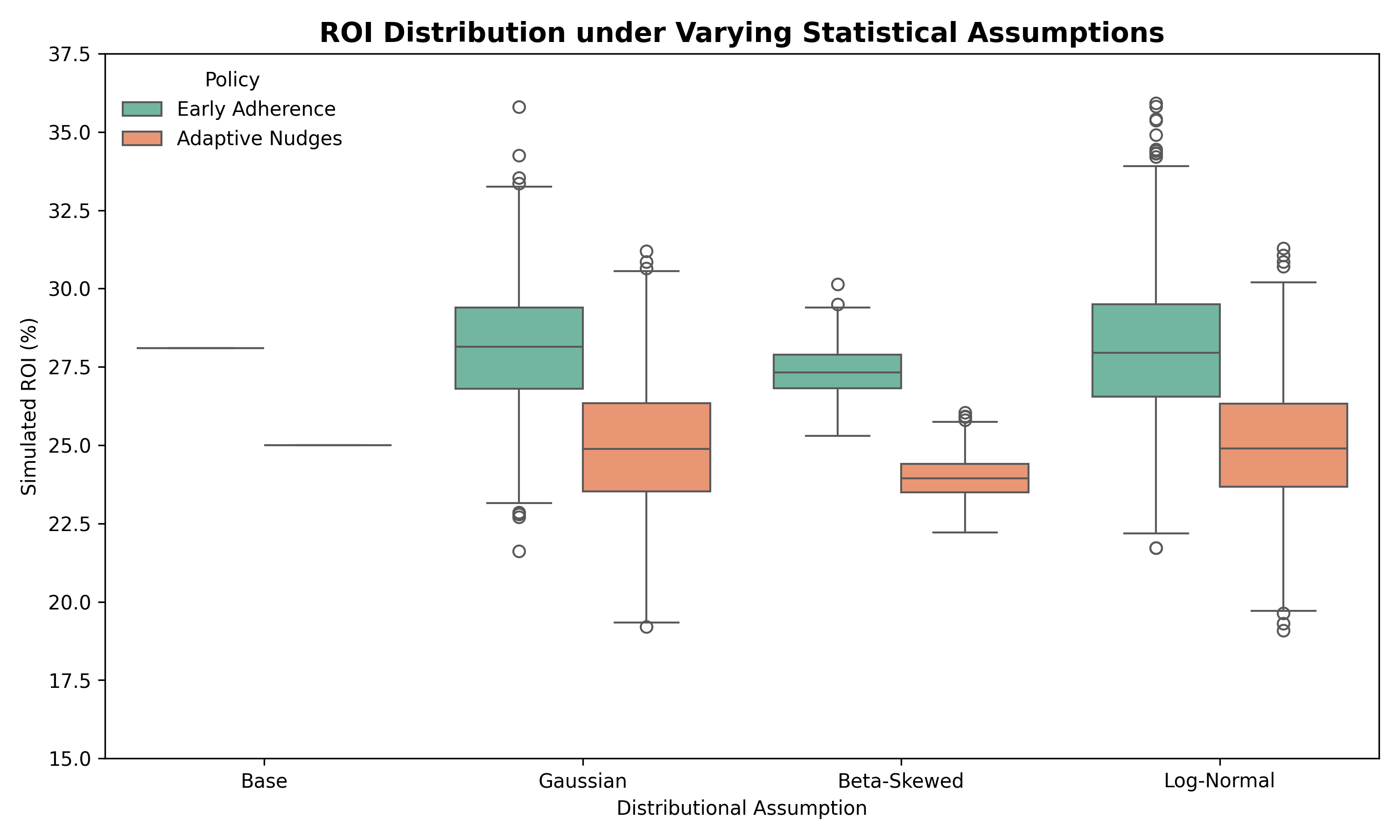}
    \caption{Mean ROI under alternative distributional assumptions. Bar heights reflect average ROI across 1,000 simulations per scenario. The comparison confirms consistent policy ordering across varying statistical structures.}
    \label{fig:roi_distributional_means}
\end{figure}

\textbf{Interpretation.}
Across a wide range of distributional assumptions, the relative ROI ranking between policy alternatives remains consistent. However, ROI magnitudes show modest sensitivity, particularly under skewed or high-variance conditions. These findings support the robustness of the proposed evaluation framework and emphasize the need for accounting for real-world variability in behavioral modeling.
\clearpage

\section{Parameter Summary}
\addcontentsline{toc}{section}{Appendix C. Parameter Summary}
\label{appendix:parameter_summary}

This appendix summarizes the key parameters and notation used throughout the simulation framework and economic evaluation model. Each symbol is defined along with its interpretation and default or tested values where applicable.

\begin{table}[htbp] 
\centering
\scriptsize
\renewcommand{\arraystretch}{1.15}
\caption{Summary of notation and parameters.}
\label{tab:parameter_summary}
% Column Definition: Changed from llp{6.5cm}l to llp{5.5cm}p{3.5cm}
\begin{tabular*}{\textwidth}{@{\extracolsep{\fill}}llp{4.5cm}p{3cm}@{}} 
\toprule
\textbf{Symbol} & \textbf{Units} & \textbf{Description} & \textbf{Value / Range} \\
\midrule
$\delta$ & --- & Magnitude of adherence improvement induced by the policy & 0.05–0.40 \\
$\gamma$ & USD (scaling factor) & Unit policy cost multiplier per patient & 0.5–1.5 \\
$C_{\text{baseline}}$ & USD & Cumulative 10-year discounted cost under no intervention & $\sim\$4{,}200$ (default) \\
$C_{\text{policy}}(\delta,\gamma)$ & USD & Cumulative 10-year discounted cost under given policy design & Scenario-dependent \\
$\text{ROI}(\delta,\gamma)$ & \% & Return on investment relative to baseline \\
$A(s)$ & [0,1] & Adherence level at time $s$ & Policy-dependent function \\
$A_0$ & [0,1] & Baseline adherence prior to intervention & 0.2 \\
$D(s;A(s))$ & [0,1] & Disease severity progression at time $s$ as a function of adherence & Logistic growth \\
$D_{\max}$ & --- & Maximum normalized disease severity & 1.0 \\
$k$ & --- & Steepness of disease progression curve & 0.5 \\
$s_0$ & Years & Midpoint of disease progression & 5.0 \\
$\alpha$ & USD & Scaling factor for disease-related cost & 600 \\
$\beta$ & USD & Coefficient for adherence-related cost term ($\beta A(s)^2$) & –1200 \\
$\rho$ & \% & Continuous annual discount rate & 3\% \\
$N$ & --- & Number of Monte Carlo simulation runs & 5{,}000–10{,}000 \\
$\mathcal{N}(\mu,\sigma^2)$ & --- & Normal distribution for stochastic adherence response & $\sigma=0.03$ \\
$\text{Beta}(\alpha,\beta)$ & --- & Beta distribution for skewed adherence scenarios & $\alpha=2$, $\beta=5$ \\
\bottomrule
\end{tabular*}
\end{table}

\textbf{Note.}  
Default parameter values are based on literature calibration and internal sensitivity analysis. Scenario-specific modifications are described in Sections~3.2 and 4.

\clearpage

%%=================================================
%%=================================================
%% Supplementary Appendix E. Model Equations
%% (Additional file 1)
%% Optional: prefix numbering for supplementary items
\renewcommand{\theequation}{SA.\arabic{equation}}
\renewcommand{\thefigure}{S\arabic{figure}}
\renewcommand{\thetable}{S\arabic{table}}
\setcounter{equation}{0}\setcounter{figure}{0}\setcounter{table}{0}
%===============================
\section{Model Equations}
\label{appendix:model_equations}
%===============================

This appendix summarizes the mathematical structure used in the simulation model.
Continuous-time notation is adopted for clarity, with discrete annual equivalents for implementation.
All costs are measured per enrollee over a 10-year horizon.

%----------------------------------------------
\subsection{Notation and Units}
\begin{itemize}
  \item $t,s$: time (years); horizon $T=10$.
  \item $\rho$: continuous discount rate; $r\approx e^{\rho}-1$ annualized rate.
  \item $C_0$: initial cost; $C(t)$: cumulative discounted cost.
  \item $c(s)$: instantaneous cost at time $s$.
  \item $D(s)$: disease burden index (increasing in severity).
  \item $A(s)\in[0,1]$: adherence level.
  \item $P(s)\in\{0,1\}$: policy activation indicator.
  \item $\beta$: adherence cost weight (can be $<0$ for savings).
  \item $\gamma$: policy cost scale.
  \item $\lambda$: optional health-outcome weight for $H(s)$ (set to $0$ in main analyses).
\end{itemize}

%----------------------------------------------
\subsection{Cumulative Cost Framework}
\begin{equation}
C(t)
= C_0 + \int_0^t e^{-\rho s}\,c(s)\,ds,
\qquad
c(s)
= D(s) + \beta A(s)^2 + \gamma P(s) + \lambda H(s).
\label{eq:SA_ct_def}
\end{equation}

The discrete-time implementation with annual steps $s=0,1,\ldots,T$ uses
\begin{equation}
C(T)
= C_0 + \sum_{s=0}^{T}\big[c_{\text{base}}(s)+\gamma P(s)+\lambda H(s)\big](1+r)^{-s},
\label{eq:SA_dt_def}
\end{equation}
where $c_{\text{base}}(s)$ aggregates the disease and adherence components.

%----------------------------------------------
\subsection{Disease and Adherence Dynamics}
Disease severity is assumed to worsen when adherence falls:
\begin{equation}
\frac{dD}{ds} = \eta[1-A(s)], \qquad \eta>0,
\label{eq:SA_D_ode}
\end{equation}
yielding \(D(s+\Delta t)=D(s)+\eta[1-A(s)]\Delta t\) in discrete time.  
Representative adherence trajectories after policy activation ($s\ge\tau$) are:
\[
A(s)=
\begin{cases}
A_0+\delta, & \text{(Early adherence; step change)}\\[3pt]
A_0+\delta e^{-\kappa(s-\tau)}, & \text{(Regressive adherence; decay rate $\kappa$)}\\[3pt]
A_0+\delta\sin(\omega s+\phi), & \text{(Adaptive nudges; periodic re-engagement)}
\end{cases}
\]
where $\delta$ represents the short-term adherence gain.

%----------------------------------------------
\subsection{Return on Investment (ROI)}
\begin{equation}
\mathrm{ROI}(\delta,\gamma)
=\frac{C_{\text{baseline}}-C_{\text{policy}}(\delta,\gamma)}
       {C_{\text{policy}}(\delta,\gamma)},
\label{eq:SA_ROI}
\end{equation}
with break-even at ROI$=0$ and a twofold return at ROI$=1$.
Cumulative present-value (PV) savings are
\(S(t)=\int_0^t e^{-\rho s}[c_{\text{base}}(s)-c(s)]\,ds\),
and the payback time \(t^\ast=\inf\{t:S(t)\ge0\}\).

%----------------------------------------------
\subsection{Scenario-Specific Instantaneous Cost Functions}

Each policy scenario Section 3 of the main manuscript 
(\textit{Intervention Scenarios}), applies the unified structure of Equation~\eqref{eq:SA_ct_def} with
scenario-specific forms of $A(s)$ and $P(s)$ as follows.

\paragraph{Scenario 1: Baseline (No Intervention).}
\begin{equation}
c_{\text{base}}(s)
=\frac{\alpha D_{\max}}{1+e^{-k(s-s_0)}}+\beta A_0^2.
\label{eq:baseline_cost_app}
\end{equation}

\paragraph{Scenario 2: Early Adherence Activation.}
\begin{equation}
c_{\text{early}}(s)
=\frac{\alpha D_{\max}}{1+e^{-k(s-s_0)}}
+\beta(A_0+\delta I(s-2))^2
+\gamma I(s-2).
\label{eq:early_cost_app}
\end{equation}

\paragraph{Scenario 3: Delayed Intervention.}
\begin{equation}
c_{\text{delayed}}(s)
=\frac{\alpha D_{\max}}{1+e^{-k(s-s_0)}}
+\beta(A_0+\delta I(s-5))^2
+\gamma I(s-5).
\label{eq:delayed_cost_app}
\end{equation}

\paragraph{Scenario 4: Regressive Adherence.}
\begin{equation}
c_{\text{regressive}}(s)
=\frac{\alpha D_{\max}}{1+e^{-k(s-s_0)}}
+\beta\big(A_0+\delta e^{-\theta(s-\tau)}I(s-\tau)\big)^2
+\gamma I(s-\tau).
\label{eq:regressive_cost_app}
\end{equation}

\paragraph{Scenario 5: Adaptive Nudges.}
\begin{equation}
c_{\text{adaptive}}(s)
=\frac{\alpha D_{\max}}{1+e^{-k(s-s_0)}}
+\beta A(s)^2
+\gamma\sum_{t\in\mathcal{T}(s)}P(t),
\label{eq:adaptive_cost_app}
\end{equation}
where $\mathcal{T}(s)$ denotes all past re-engagement times.

\paragraph{Scenario 6: High-Cost, Low-Impact Policy.}
\begin{equation}
c_{\text{lowimpact}}(s)
=\frac{\alpha D_{\max}}{1+e^{-k(s-s_0)}}
+\beta(A_0+\delta)^2
+\gamma P(s),
\qquad \gamma\gg1,\;\delta\ll1.
\label{eq:lowimpact_cost_app}
\end{equation}

%----------------------------------------------
\subsection{Decomposition for Intuition}
For interpretive clarity,
\begin{equation}
C_{\text{policy}}(\delta,\gamma)
= C_0 - f(\delta) + g(\gamma,\delta),
\label{eq:SA_policy_decomp}
\end{equation}
where \(f(\delta)\) captures adherence-driven savings and \(g(\gamma,\delta)\) denotes implementation cost.
Increasing $\delta$ or decreasing $\gamma$ both improve ROI as in Equation~\eqref{eq:SA_ROI}.